\documentclass[sn-mathphys-num]{sn-jnl}

\usepackage{xcolor}
\usepackage[normalem]{ulem}

\definecolor{darkblue}{rgb}{0.,0.,0.75}
\definecolor{darkred}{rgb}{0.5,0.,0.}
\definecolor{darkgreen}{rgb}{0.0,0.5,0.}
\usepackage{amsmath, amssymb,mathtools,amsthm,amsfonts,amssymb,braket,comment,mathtools,bbm}
\usepackage[all]{xy}
\usepackage{fullpage}
\usepackage{chemarrow}
\usepackage{marginnote}
\usepackage[numbers]{natbib}
\usepackage{tikz-cd}
\usepackage{scalerel}
\newtheorem{lemma}{Lemma}[section]

\newtheorem{proposition}[lemma]{Proposition}
\newtheorem{proposition-definition}[lemma]{proposition-Definition}
\theoremstyle{definition}
\newtheorem{definition}[lemma]{Definition}
\newtheorem{remark}[lemma]{Remark}
\newtheorem{example}[lemma]{Example}
\usepackage{enumitem}

\usepackage{appendix}
\numberwithin{equation}{section}
\theoremstyle{definition}

\makeatletter 
\def\thm@space@setup{%
  \thm@preskip=6pt
  \thm@postskip=6pt
}
\makeatother

\def\CA{{\mathcal A}}

\def\CM {{\mathcal M}}

\def\CD {{\mathcal D}}

\def\CW {{\mathcal W}}
\def\CX {{\mathcal X}}

\def\CZ {{\mathcal Z}}

\def\CG {{\mathcal G}}
\def\CH {{\mathcal H}}

\def\CD {{\mathcal D}}
\def\CQ {{\mathcal Q}}
\def\CS {{\mathcal S}}

\def\CX {{\mathcal X}}

\def\CZ{{\mathcal Z}}

\def\pd{\partial}
\def\h{{\mbox{H}}}
\def\L{{\mbox{L}}}
\def\hm{{\mbox{H}}^{-}}

\def\ep{\varepsilon}

\def\IF{\mathbb{F}}

\def\IC{\mathbb{C}}
\def\IQ{\mathbb{Q}}
\def\IR{{\mathbb{R}}}

\def\IZ{{\mathbb{Z}}}

\def\bZ{\mathbb{Z}}

\def\la{\langle}
\def\a{\alpha}

\def\ra{\rangle}

\def\End{\operatorname{End}}
\def\La{{\Lambda}}
\def\l{{\lambda}}
\def\hl{{\hat \lambda}}

\def\sc{\scaleto{\,\circ\,}{4pt}}
\def\sl{\mathcal{S}_{\textsuperscript{L}}}

\def\Im{\operatorname{im}}
\def\ker{\operatorname{ker}}
\def\Hom{\operatorname{Hom}}
\def\coker{\operatorname{coker}}
\def\lr{\longrightarrow}

\def\ext{\operatorname{Ext}}
\def\tr{\operatorname{tr}}
\def\lab{\operatorname{lab}}

\newcommand{\id}{\mathbbm{1}}

\newcommand{\dd}{{\mathsf{d}}} 

\begin{document}

\title[Invertible Stabilizer Codes]{A Classification of Invertible Stabilizer Codes}

\author*[1]{\fnm{Roman} \sur{Geiko}}\email{geiko@physics.ucla.edu}

\author*[2]{\fnm{Georgii} \sur{Shuklin}}\email{g.shuklin@sheffield.ac.uk}

\affil*[1]{\orgdiv{Department of Physics and Astronomy}, \orgname{University of California}, \orgaddress{\city{Los Angeles}, \postcode{90095}, \state{CA}, \country{USA}}}

\affil[2]{\orgdiv{School of Mathematical and Physical Sciences}, \orgname{University of Sheffield}, \orgaddress{\country{UK}}}

\abstract{We develop a framework for the classification of invertible translation‑invariant stabilizer codes modulo condensation and stabilization with simple codes. Our stabilizer codes generalize the Pauli codes: they are extracted from local commuting projector Hamiltonians whose terms are made of local alphabets more general than the Pauli operators. In particular, not all such codes can be entangled/disentangled with Clifford Quantum Cellular Automata. We compute their Witt-equivalence classes in any spatial dimension in terms of relative L‑theory groups. In particular, we show that the group of equivalence classes of such codes in three spatial dimensions is isomorphic to the Witt group of abelian topological orders in two spatial dimensions. In four spatial dimensions we obtain a group of order two, conjecturally corresponding to the non-trivial invertible phase.  Additionally, we propose a representing spectrum for the generalized cohomology theory corresponding to the invertible stabilizer codes.}


\maketitle

\section{Introduction}\label{sec:intro}
Entanglement patterns in the ground states of quantum lattice systems play a central role in contemporary quantum many-body physics. The least entangled classes of many-body states, beyond product states, are the invertible states introduced by A. Kitaev \cite{Kitaev}. Despite their relative simplicity, a complete and rigorous theory of them still does not exist. The group of invertible phases is known only in one spatial dimension \cite{Kapustin_2021}, though there are many conjectures coming from the effective field theory descriptions \cite{kapustin2014symmetryprotectedtopologicalphases,xiong2018minimalist,Gaiotto_2019,Freed_2021,Shiozaki2022,kubota2025stablehomotopytheoryinvertible}. Here, we introduce \textit{invertible stabilizer codes} arising from a class of Hamiltonians with finite-range interactions. The algebraic nature of such codes allows us to compute their equivalence classes in any spatial dimension. After quotienting the classes induced from lower-dimensional systems,
the resulting reduced groups are labeled by the spatial dimension
$\dd$ and are given below
\begin{align}\label{eq:phases}
\operatorname{ISC}_\dd=\begin{cases}\CW^{pt}\,,\;\;\,\, \dd\equiv 3\;\,\mbox{mod}\;4\; \&\; \dd>0\,,\\
  \IZ/2\,,\quad \dd\equiv 0\;\,\mbox{mod}\;4\; \&\; \dd>0\,,\\
  0\,,\quad\quad\; \mbox{otherwise}\,.\\
  \end{cases}
\end{align}
Here, $\CW^{pt}$ is the Witt group of linking (torsion) quadratic forms on finite abelian groups, see \eqref{eq:witt_groups_quad_forms}, also known as the Witt group of pointed modular tensor categories, also known as the Witt group of two-dimensional abelian topological orders. 

Our methods are inspired by the symplectic formalism which, in its original form, is applicable to the Pauli stabilizer codes \cite{haah2013,haah2016,haah2021classification,haah2022topological,haah2021clifford,ruba2024homological,Geiko_2024}, i.e., the ground states of Pauli stabilizer Hamiltonians. We extend the symplectic formalism by replacing the Pauli groups with more general twisted quasi-Pauli groups, which are essentially groups of Weyl operators constructed by an anti-hermitian form on a certain class of modules over the ring  of Laurent polynomials with integer coefficients. Then, our algebraic stabilizer codes are determined by finitely-generated isotropic $R$-submodules of the modules of labels of the Weyl operators. The process of condensation allows us to pass between non-isomorphic stabilizer submodules and different quasi-Pauli groups, in contrast with the relation allowed for the Pauli codes. We call a locally topologically ordered code invertible when its topological-charge groups vanish. The key technical point is the reformulation of invertible stabilizer codes in terms of the linking formations. Then, the Witt group of formations corresponds to the group of stabilizer codes modulo stabilization and condensation. 

We expect that the equivalence classes of invertible stabilizer codes coincide with the classes of QCA modulo finite-depth quantum circuits and translations that entangle/disentangle the corresponding codes. We leave a rigorous derivation of that connection for the future while outlining a relation between the invertible stabilizer codes and locally-flippable separators from \cite{haah2023nontrivial}, which do correspond to QCA. Indeed, invertible stabilizer Hamiltonians were used in \cite{haah2023nontrivial, haah2021clifford, fidkowski2022gravitational,Chen_2023} to argue that the QCA disentangling them are non-trivial as long as the Hamiltonians are non-trivial. If we assume that the classes of codes are in one-to-one correspondence with the classes of QCA, then our results for $\dd=3$ confirm the results of \cite{shirley2022three}, for $\dd=4$, our results confirm \cite{fidkowski2022gravitational,Chen_2023}, and for $\dd=5$, they confirm \cite{Fidkowski_2025}. The appearance of the Witt classes of topological orders \cite{Johnson_Freyd_2022} in the classifications of QCA was explained in \cite{haah2021clifford}. As was mentioned above, $\operatorname{ISC}_3$ is isomorphic to the Witt group of \textit{abelian} topological orders in $\dd=2$, while $\operatorname{ISC}_4$ and $\operatorname{ISC}_5$ are isomorphic to the Witt groups of topological orders in $\dd=3$ and $\dd=4$, respectively. The Witt group of topological orders in $\dd=3$ was conjectured to be $\IZ/2$ in \cite{johnsonfreyd2020Z2} and this result was proven in \cite{Johnson_Freyd_2023}. 

Remarkably, we obtain a classification of invertible stabilizer codes using $L$-theory, extending methods previously applied to Clifford QCA and Pauli codes. Topological $K$-theory provides standard classification tools for free-fermion phases \cite{Kitaev_2009,freed2013twisted}, while algebraic $L$-theory gives a handle on the interacting stabilizer setting considered here.The groups $\operatorname{ISC}_{\dd}$ in \eqref{eq:phases} are the relative $L$-groups, the groups participating in the long exact sequence of homotopy groups of a cofiber of a map of spectra, see \S \ref{sec:Lspectra}. We prove that this particular spectrum is a generalized Eilenberg-MacLane spectrum. 
\subsection*{Organization of the paper}
We review $(R,S)$-modules, discuss their $K$-theory, and characterize them in terms of topological charges in \S \ref{sec:RS_mods}. We define linking forms and formations in \S \ref{sec:linking forms}. We then introduce Pauli, quasi-Pauli, and twisted quasi-Pauli groups, together with the algebra of local operators generated by them, in \S \ref{sec:unitary_groups}. In \S \ref{ref:classification} we define stabilizer Hamiltonians and invertible stabilizer codes and identify them with non-singular linking formations. Finally, we calculate the relevant relative $L$-groups in \S \ref{sec:L-groups} and describe the corresponding $L$-theory spectrum in \S \ref{sec:Lspectra}.

\section{$(R,S)$-modules}\label{sec:RS_mods}
 In this section, we review relevant aspects of the theory of torsion modules. We will denote by $R$ the ring of Laurent polynomials with integer coefficients $R\coloneqq \IZ[\La]=\IZ[x_1,x_1^{-1},\ldots, x_{\dd},x_{\dd}^{-1}]$. By $S$ we will denote a multiplicatively closed subset $S\coloneqq\IZ-\{0\}\hookrightarrow R$  embedded as the constant polynomials.\footnote{ An attentive reader might be able to split the results that apply only to the ring of Laurent polynomials from the results that would be valid for any commutative ring with identity and involution.}
 We view $R$ as a ring with the following involution
\begin{align}
   \bar x_1=x_1^{-1}\,,\ldots,\, \bar x_{\dd}=x_{\dd}^{-1}\,
\end{align}
which does not act on the coefficients of the Laurent polynomials. 
 Localization of $R$ with respect to $S$ is denoted by $S^{-1}R$ and localization of an $R$-module $M$ is $S^{-1}M=M\otimes_R S^{-1}R$. In our case, $S^{-1}R=\IQ[\La]$. An $R$-module $M$ is called $S$-torsion if $S^{-1}M=0$. The central role in this work is played by the torsion $R$-modules of homological dimension one. 
 
\begin{definition}[\cite{Ranicki1981ExactSI}, p.181]\label{def:RSmodule}
    An $(R,S)$-module is a finitely-generated $S$-torsion $R$-module of homological dimension one, i.e., it admits a projective resolution of length one.
\end{definition}
 \vspace{3mm}
 
 Throughout this work, we will implicitly use the Serre-Suslin-Swan theorem stating that all FGP modules over $R$ are free \cite{Suslin1977,Swan1978}  . In particular,  for an $(R,S)$-module, we can only consider the resolutions that have the form 
 \begin{align}\label{eq:liftoftorsionmod}
        0\rightarrow R^n \xrightarrow{\pd} R^n\xrightarrow{} M\rightarrow 0,
    \end{align}
     where $\pd$ becomes an isomorphism after inverting all elements of $S$.

 We notice that all $(R,S)$-modules with $\dd=0$ are isomorphic to some finite abelian group $\CD$. We can always bring the matrix of $\pd$ to its Smith normal form, which has positive integers on the diagonal. Then, for $\dd=0$, any $(R,S)$-module $M=\coker \pd$, is given by the corresponding finite abelian group.

For $\dd>0$, $(R,S)$-modules are more diverse. One simple class is given by the modules of the form $\CD\otimes_\IZ R$ where $\CD$ is some finite abelian group, however not all $(R,S)$-modules have this form.
\vspace{3mm}

\begin{remark}\label{the_remark_about_det(pd)} Note that the free resolution (\ref{eq:liftoftorsionmod}) is not unique since we can pre- and post-compose $\pd$ with automorphisms of $R^n$. Since $\pd$ becomes invertible over $S^{-1}R=\IQ[\La]$, one has \begin{align} \det\pd=a x^u, \qquad a\in\IZ\setminus\{0\}, \quad u\in\La. \end{align} Pre- or post-composing with an automorphism of determinant $\pm x^{-u}$ allows us to arrange that $\det\pd=|a|$ is a positive integer. Throughout the paper we use such normalized resolutions.
\end{remark}

\vspace{3mm}

\begin{remark} \label{remark:ktheory}(\textbf{K-theory of $(R,S)$-modules}). Let $T$ be the category of $(R,S)$-modules. This category is exact because it is closed with respect to taking extensions: an extension of an $(R,S)$-module by another $(R,S)$-module is an $(R,S)$-module, which follows from the Horseshoe Lemma. The Grothendieck group $K_0(T)$ \cite{QuillenHigherK} is generated by the classes $[M]$ subject to the relation $[M]=[M_1]+[M_2]$ whenever $M$ is an extension of $M_1$ by $M_2$.

We calculate  $K_0(T)$ by using the localization exact sequence of Quillen \cite{Grayson}. Let us denote by $K_n(R)$ the $K$-groups of the category of FGP modules over $R$. Then, we have an exact sequence (Theorem 7.1, \cite{Grayson}):
\begin{align}\label{eq:localizationKtheory}
  \ldots \rightarrow  K_1(T) \rightarrow K_1(R)\rightarrow K_1(S^{-1}R)\rightarrow 
  K_0(T) \rightarrow K_0(R)\rightarrow K_0(S^{-1}R)\;.
\end{align}

A simple computation using the fundamental theorem of $K$-theory \cite{KBook} leads to an isomorphism
\begin{align}\label{eq:detmap}
 K_0(T)= \coker (\IZ/2\to \IQ^{\times})= \IQ^{\times}_+   
\end{align}
 where $\IQ^{\times}_+$ is the multiplicative group of positive rational numbers. We can describe the isomorphism \eqref{eq:detmap} explicitly. By the previous remark, any $(R,S)$-module $M$ can be associated with a positive integer $\det \pd$, such that the class of $M$ in $K_0(T)$ is given by $\det \pd$. 
\end{remark}

\subsection{Examples}\label{sec:examples} Let us consider a class of examples of $(R,S)$-modules given by the generating matrix
\begin{align}\label{(2.3)}
    \pd=\begin{pmatrix}
        p & f\\
        0 & p
    \end{pmatrix}\,,\quad f\in \IZ[\Lambda]\,.
\end{align}

\begin{example}\label{example:f_inv}

Let $f=x$. Then $M=\coker\pd$ is isomorphic to $R/p^2R$. Indeed, it is defined by generators $e_1,e_2$ and relations \begin{align} pe_1=0, \qquad xe_1+pe_2=0. \end{align} Since $x$ is a unit, we can eliminate $e_1=-px^{-1}e_2$, and the remaining relation is $p^2e_2=0$. Thus $e_2$ generates $M$ and $M\cong R/p^2R$.

For any prime $p$, we use the Horseshoe Lemma (Lemma 2.2.8, \cite{weibel1994introduction}), to demonstrate that the  matrix \eqref{(2.3)} comes from a certain extension  of $\mathbb{Z}/p[\Lambda]$ by $\mathbb{Z}/p[\Lambda]$.   Let us calculate $\text{Ext}^1_{\mathbb{Z}[\Lambda]}(\mathbb{Z}/p[\Lambda], \mathbb{Z}/p[\Lambda])$. Using $\text{Hom}(- , \mathbb{Z}/p[\Lambda])$ and the resolution $\mathbb{Z}[\Lambda] \xrightarrow{p} \mathbb{Z}[\Lambda]$, we obtain 
\begin{align}
    \text{Ext}^1_{\mathbb{Z}[\Lambda]}(\mathbb{Z}/p[\Lambda], \mathbb{Z}/p[\Lambda]) = \mathbb{Z}/p[\Lambda]. 
    \end{align} 
    
It follows that two matrices with $f=f_1$ and $f=f_2$ such that $f_1-f_2$ is divisible by $p$ give isomorphic modules. 

 On the other hand,  the multiplicative group $(\mathbb{Z}[\Lambda])^{\times}$ acts on  $ \text{Ext}^1_{\mathbb{Z}[\Lambda]}(\mathbb{Z}/p[\Lambda], \mathbb{Z}/p[\Lambda])$ and two extensions of $\mathbb{Z}/p[\Lambda]$ by $\mathbb{Z}/p[\Lambda]$  define isomorphic modules if they lie on the same orbit. So,  $f=1$ and $f=x $ give the same module $\IZ/p^2[\Lambda]$.

\end{example}
 \vspace{3mm}
\begin{example}\label{Example_RS_mod_non_invf}
Let $\dd=1$, $p=2$, and $f=x+x^{-1}$. Let us show that in this example $M$ is not isomorphic to $\CD\otimes_{\IZ}R$ for any finite abelian group $\CD$.
Suppose the opposite: $M \cong \CD \otimes_{\IZ} \mathbb{Z}[x, x^{-1}]$. Note that the isomorphism $\mathbb Z\cong \mathbb{Z}[x, x^{-1}] / (x-1) $ induces a structure of  $R$-module on $\mathbb Z$.  Then, 
\begin{align}
 M \otimes_{\mathbb{Z}[x,x^{-1}]} \mathbb{Z} \cong \CD \otimes_{\mathbb{Z}} \mathbb{Z}[x, x^{-1}] \otimes_{\mathbb{Z}[x, x^{-1}]} \mathbb{Z} \cong  \CD
\end{align}
where $\IZ$ is viewed as a $\IZ[x,x^{-1}]$-module via the evaluation map.

The module $\CD\cong M\otimes_{\mathbb{Z}[x,x^{-1}]} \mathbb{Z}[x,x^{-1}]/(x-1)$ is defined by the matrix $\begin{pmatrix}2&2\\0&2\end{pmatrix}$ and is isomorphic to $\mathbb Z/2\oplus\mathbb Z/2$. Therefore, we would have $M\cong(R/2R)^2$, in which every element has order at most $2$. On the other hand, the image of $e_1$ generates a submodule isomorphic to $R/2R$: if $re_1=0$, then $(r,0)=\pd(a,b)$ for some $a,b\in R$; the second component gives $b=0$, and hence $r=2a$. Thus $(x+x^{-1})e_1\neq0$. Since \begin{align} 2e_2=-(x+x^{-1})e_1, \end{align} the element $e_2$ does not have order $2$, a contradiction.
\end{example}
\vspace{3mm}

\begin{remark}\label{phys_ex}
We can interpret the $(R,S)$-module from Example \ref{Example_RS_mod_non_invf} as giving rise to a constrained Hilbert space. First, we notice that $M\cong M\otimes_R R/4$ and we can obtain it as a cokernel of the generated map reduced modulo $4$:
\begin{align}
    (R/4)^{\oplus 2}\xlongrightarrow{\pd=\left(\begin{smallmatrix}
        2 & x+x^{-1}\\0 &2
    \end{smallmatrix}\right)} (R/4)^{\oplus 2}\longrightarrow M'\longrightarrow 0 \,.
\end{align}
 This reduction allows us to obtain $M$ from imposing constraints on $(R/4)^{\oplus 2}$ which is countable as a set. This procedure has the following physical interpretation. We start with a one-dimensional lattice $\IZ$ with two four-dimensional degrees of freedom on each site. Each site is acted by the $\IZ/4$ Pauli operators, see \S \ref{Sec:Pauligroup} and \S \ref{sec:lattice_pauli_group}, that we label by $X_j^A, X_j^B$ and $Z_k^A, Z_k^B$ with $j,k\in \IZ$ and $(X_j^A)^4=(X_j^B)^4=1$, $(Z_k^A)^4=(Z_k^B)^4=1$, where $A$ and $B$ label degrees of freedom within each site. Next, we impose the constraints: 
\begin{align}\label{eq:constraint}
    X_{j-1}^B(X_{j}^A)^2X_{j+1}^B=1\,\;\&\; (X_j^B)^2=1\,\quad  \mbox{for all}\; j\in \IZ\,.
\end{align}
In other words, the constrained Hilbert space is obtained from $\ell^2((R/4)^{\oplus 2})$ by a two-step reduction to the ground spaces of the Hamiltonians $H_1$ and $H_2$: 
\begin{align}
    H_1=\sum_{i\in \IZ}\left(\frac{1-(X^B_i)^2}{2}\right)\,,\quad H_2=\sum_{i\in \IZ}\left(\frac{1-(X^B_{i-1})(X_i^A)^2X_{i+1}^B}{2}\right)\,.
\end{align}
    Remarkably, the new operators acting on the constrained Hilbert space are not given by the ordinary Pauli operators from \S \ref{sec:lattice_pauli_group} but rather by the twisted quasi-Pauli operators from \S \ref{sec:twisted_quasi_pauli}.
    \end{remark}
\vspace{3mm} 

\subsection{Compactification} It is a common approach in physics to compactify the lattice $\La$ to a torus with a finite number of elements and impose periodic boundary conditions. Motivated by that application, we call the following map \textit{compactification}:
\begin{align}
   R\to R_{\tau}\coloneqq R/\mathfrak{b}_{\tau}\,.
\end{align}
where $\mathfrak{b}_{\tau}$ is an ideal in $R$ generated by $\{x_1^{\tau}-1,\ldots, x_{\dd}^{\tau}-1\}$.  In view of Remark \ref{the_remark_about_det(pd)}, we have the following proposition.
\begin{proposition}\label{prop_counting}
      Let $M=\coker \pd$ be an $(R,S)$-module and let $M_{\tau}=M\otimes_{R}R_{\tau}$. Then, the number of elements $|M_{\tau}|$ in $M_{\tau}$ is given by
      \begin{align}
          |M_{\tau}|=(\det \pd )^{\tau^\dd}
      \end{align}
\end{proposition}
\begin{proof}
Let $\mathrm{Res_{\bZ}}:R_{\tau}\!-\!\mathrm{Mod \to \bZ\!-\!Mod }$ be the restriction functor. Since $M$ is an $(R,S)$-module, there is $a\in \mathbb N$ such that $aM=0$ and $a\mathrm{Res_\bZ}(M_{\tau})=0$. On the other hand, both functors $-\otimes_R R_{\tau}$ and $\rm Res_\bZ$ map f.g. modules to f.g. modules. Thus, $\mathrm{Res_\bZ}(M_{\tau})$ is a finite abelian group and the number of elements can be computed as $|\mathrm{det}A|$ where $\bZ^{N}\overset{A}{\longrightarrow} \bZ^{N}$ is a resolution of $\mathrm{Res_\bZ}(M_{\tau})$.

Now, let $R^n\overset{\pd}{\longrightarrow} R^n$ be a resolution of $M$. Note that the evaluation functor $-\otimes_R R_{\tau}$ is right-exact. So $M_{\tau}\simeq \mathrm{ coker([\pd])}$ where $[\pd]$ is the matrix with the entries $\pd_{ij} {\rm mod}\, \mathfrak{b}_{\tau}$. Then $\mathrm {Res_\bZ([\pd])}$ is the $n\times n$ block-matrix with $\tau^\dd\times \tau^\dd$  blocks $D_{ij}$. Since $R_\tau$ is a commutative ring, the matrices $D_{ij}$ mutually commute. Now, we need to compute the determinant of a block-matrix with commuting blocks. By Theorem 1 in \cite{kovacs}, we have:
\begin{align}
\det (Res_\bZ([\pd]))=\det (\Sigma_{\sigma \in S_n} (-1)^{|\sigma|} D_{1\sigma(1)}\ldots   D_{n\sigma(n)}).
\end{align}
It remains to note that $\rm Res_\bZ$ is additive so that 
$\mathrm{Res_\bZ:End}_{R_{\tau}}(R)\to  \mathrm{End_{\bZ}}(R^{\tau^\dd})$
is  a homomorphism of algebras. Then, we have
\begin{align}
    \Sigma_{\sigma \in S_n}  (-1)^{|\sigma|} D_{1\sigma(1)}\ldots  D_{n\sigma(n)}= \mathrm{Res_\bZ(det[\pd])}.
\end{align}
Finally, since $\det\pd$ is an integer, $\rm det[\pd]$ is the same integer,  ${\rm Res_\bZ(det[\pd])=det \pd} \cdot E$, where $E$ is the identity $\tau^\dd\times \tau^\dd$ matrix, and $\det \mathrm{Res}_\bZ([\pd]) = (\det \pd)^{\tau^\dd}$. 
\end{proof} 

\subsection{Topological charges}\label{sec:top_charges}

This section is devoted to obtaining an auxiliary result characterizing  $(R,S)$-modules in terms of their $\ext$-groups. For reasons explained in \ref{sec:codes/lagrangians}, we call the $\ext$-groups participating in the following Proposition ``topological charges". 

\begin{proposition}\label{prop:top_charges}
    Let $M$ be a finitely-generated $S$-torsion $R$-module. Then, the following three conditions are equivalent: 
    \begin{enumerate}
    \item The projective dimension of $M$ is less than or equal to $1$.
    \item $\ext_{R}^i(M,S^{-1}R/R)=0\,,\quad i\geqslant1\,.$
        \item Let $n\in S$ be such that $nM=0$, then $\ext_{R/nR}^i(M,R/nR)=0$, $i\geqslant 1$.
    \end{enumerate}
\end{proposition}
\begin{proof}
$(1)\Leftrightarrow (2)$ The projective dimension of $M$ is less than or equal to $1$ if and only if $\ext^i_R(M,P)=0$ for all $i\geqslant 2$ and any $R$-module $P$. Since $R$ is regular Noetherian of finite global dimension, it is also equivalent to the condition $\ext^i_R(M,R)=0$ for all $i\geqslant 2$.

Given the short exact sequence 
\begin{align}
    0\longrightarrow R\longrightarrow S^{-1}R\longrightarrow S^{-1}R/R\longrightarrow0\,
\end{align}
we obtain the associated long exact sequence of $\ext$-groups:
\begin{align}\label{eq:ext_long_ex_seq}
\ldots \longrightarrow\ext_R^i(M,R)\longrightarrow \ext_R^i(M,S^{-1}R)\longrightarrow \ext_R^i(M,S^{-1}R/R)\longrightarrow\ldots  
\end{align}

 Since $M$ is finitely generated, $\ext$ commutes with localization. And as $M$ is $S$-torsion, we obtain $\ext^i_R(M,S^{-1}R)=\ext^i_{S^{-1}R}(S^{-1}M,S^{-1}R)=0$ for every $i\geqslant0$. Then, using \eqref{eq:ext_long_ex_seq}, we obtain
\begin{align}
   \ext_R^{i}(M,R)=\ext_R^{i-1}(M,S^{-1}R/R)=0 \,,\quad i\geqslant 2\,. 
\end{align}

$(1)\Leftrightarrow (3)$ 
As $M$ is a finitely generated $S$-torsion module, there exists $n\in S$ such that $n M=0$. Let us use the  change-of-rings spectral sequence, see, e.g. Theorem 10.75 in \cite{rotman2009introduction}:
\begin{align} 
E_2^{p,q}
=
\operatorname{Ext}^p_{R/nR}(M, \operatorname{Ext}^q_R(R/nR,R))
\Longrightarrow
\operatorname{Ext}^{p+q}_R(M,R).
\end{align} 

A 2-term free resolution of \(R/n\) shows:
\begin{align} 
\operatorname{Ext}^0_R(R/nR,R)=0,\qquad
\operatorname{Ext}^1_R(R/nR,R)\cong R/nR,\qquad
\operatorname{Ext}^q_R(R/nR,R)=0 \ (q\ge 2).
\end{align} 
Thus only the row \( q=1 \) survives, giving for all \( p\ge 0 \):
\begin{align}\label{ext_shift}
\operatorname{Ext}^p_{R/nR}(M,R/nR)
\cong
\operatorname{Ext}^{p+1}_R(M,R).
\end{align}

Therefore, $(1)$ is equivalent to $(3)$.
\end{proof}

\section{Linking forms and formations}\label{sec:linking forms}
In this section we review the linking forms on $(R,S)$-modules. 

Let $N$ be a finitely-generated projective $R$-module. Then, its algebraic dual is $N^*=\Hom_{R}(N,R)$ with the following action of $R$:
\begin{align}
   r\times f\mapsto (a\mapsto f(a)\,\bar{r})\,,\quad r\in R,a\in N,f\in N^{*}\,.
\end{align}
The algebraic dual of a morphism $\phi\in \Hom_R(N_1,N_0)$ is a morphism $\phi^*\in \Hom_R(N_0^*,N_1^*)$ defined as 
\begin{align}
    \phi^*:f\mapsto (a\mapsto f(\phi(a)))\,.
\end{align}

The double dual of a finitely-generated projective module is naturally isomorphic to itself $N^{**}=N$. Evidently, the algebraic dual of a torsion module is zero, and for that reason, for $(R,S)$-modules we use an analogue of the Pontryagin dual.

\begin{definition}[\cite{Ranicki1981ExactSI}, p. 181]
    The $S$-dual of an $(R,S)$-module $M$ is an $(R,S)$-module  $M^{\vee}\coloneqq\Hom_R(M,S^{-1}R/R)$
with $R$ acting as follows
\begin{align}
   r\times f\mapsto (a\mapsto f(a)\,\bar{r})\,,\quad r\in R,\,a\in M,\,f\in M^{\vee}\,.
\end{align}
For example, if $\dd=0$, any $(\IZ,\IZ-\{0\})$-module is a finite abelian group $\CD$ and $\CD^{\vee}$ is its Pontryagin dual. If $M$ is resolved by \eqref{eq:liftoftorsionmod}, then $M^{\vee}$ admits a projective resolution 
    \begin{align}
        0\rightarrow (R^n)^{*} \xrightarrow{\pd^*}(R^n)^*\xrightarrow{Th}M^{\vee}\rightarrow 0
    \end{align}
where
\begin{align}
    Th:(R^n)^*\to M^{\vee}\,,\quad\quad
    f\mapsto \left([x]\to \frac{f(y)}{s}\right)
\end{align}
where $x,y\in R^n$, $[x]\in M$, $s \in S$, and $s\,x=\pd\, y$.
\end{definition}

Relative to the standard bases, the map $m\mapsto\bar m$ identifies $R^n$ with $(R^n)^*$. We therefore obtain that $M^\vee$ is generated by the matrix $\bar{\pd}^T$. Also note that there is a natural isomorphism $M^{\vee\vee}\cong M$ for any $(R,S)$-module $M$.

\subsection{Linking forms}
Throughout the paper, $\ep$ will stand for $\pm$. 

\begin{definition}[\cite{Ranicki1981ExactSI}, p.223]
    An $\ep$-hermitian linking form over $(R,S)$ is a pair $(P,\lambda)$ where $P$ is an $(R,S)$-module and $\l$ is an $R$-module map $\l:P\to P^{\vee}$ such that $\l^{\vee}=\ep \l$. Equivalently, it defines a pairing $\hat \l$:
\begin{align}
       &\hl: P\times P\rightarrow S^{-1}R/R\,,\nonumber\\
       \hl(x,ry)=r\hl(x,y)\,,\quad &\hl(x,y_1+y_2)=\hl(x,y_1)+\hl(x,y_2)\,,\quad \hl (x,y)=\ep\overline{\hl(y,x)}\,,
\end{align}
for $x,y_1,y_2\in P\,,r\in R$. The form is non-singular if $\l$ is an isomorphism. 
A morphism of $\ep$-hermitian linking forms $(P_1,\l_1)\to(P_2,\l_2)$ is a morphism  of $R$-modules $f:P_1\to P_2$ such that $\l_1=f^{\vee}\l_2f$. Given an $(R,S)$-module $M$, the \textit{standard $\ep$-hermitian form}, $\h^{\ep}(M)$ is an $(R,S)$-module $M\oplus M^{\vee}$ with the $\ep$-hermitian form $\l^{\ep}_0$ with the matrix $\big(\begin{smallmatrix}
    0 & +1\\
    \ep\,1 & 0
\end{smallmatrix}\big)$.
\end{definition} 
We will call $(+)$-hermitian forms simply hermitian, and $(-)$-hermitian forms anti-hermitian. 

\begin{example}\label{example_2}
Let $M$ be the module from Example \ref{Example_RS_mod_non_invf} with $\pd=\begin{pmatrix} 2 & x+x^{-1} \\ 0 & 2 \end{pmatrix}$.
 Then the dual module is given by the matrix
\begin{align}
    \partial^\ast = \begin{pmatrix} 2 & x+x^{-1} \\ 0 & 2 \end{pmatrix}^T = \begin{pmatrix} 2 & 0 \\ x^{-1}+x & 2 \end{pmatrix}\,. 
\end{align}
The matrix $\kappa_{12} = \begin{pmatrix} 0 & 1 \\ 1 & 0 \end{pmatrix}$  induces a commutative diagram
\begin{equation}
\begin{tikzcd}
R^2 \arrow[r, "\kappa_{12}"] \arrow[d, "\partial"] & R^2 \arrow[d, "\partial^\ast"] \\
R^2 \arrow[r, "\kappa_{12}"] & R^2
\end{tikzcd}
\end{equation}
and defines an isomorphism $\l : M \to M^\vee$. Let us calculate the corresponding form $\hl$. Let  $e_1, e_2$ be a basis for the top $R^2$, $m_1, m_2$ be a basis for the bottom $R^2$, and $e_1^\ast,e_2^\ast, m_1^\ast,m_2^\ast$ be the dual bases. Then, $\kappa_{12}$ maps
\begin{align}
 m_1 \mapsto e_2^\ast = \hl(m_1, -),\,\,\,\,\, m_2 \mapsto e_1^\ast = \hl(m_2, -)\,.
 \end{align}
Note that $4m_2 = \partial(-(x+x^{-1})e_1 + 2e_2)$ and $2m_1 = \partial e_1$. Then:
\begin{align}
&\hl(m_1, m_1) = e_2^\ast(m_1) = \frac{e_2^\ast(e_1)}{2} = 0\,,\\
&\hl(m_1, m_2) = e_2^\ast(m_2) = \frac{e_2^\ast(-(x+x^{-1})e_1 + 2e_2)}{4} = \frac{1}{2}\,,\\
&\hl(m_2, m_2) = \frac{e_1^\ast(-(x+x^{-1})e_1 + 2e_2)}{4} = -\frac{x+x^{-1}}{4}\,,\\
&\hl(m_2, m_1) = \frac{e_1^\ast(e_1)}{2} = \frac{1}{2}=-\frac{1}{2}\,\in \mathbb Q/\bZ.
\end{align}  
\end{example}
\vspace{3mm}

\begin{remark}\label{remark_commutator}
    The ring $\IZ[\La]=\IZ[x_1,x_1^{-1},\ldots, x_{\dd},x_{\dd}^{-1}]$ comes with a $\IZ$-linear map
\begin{align}
    \tr:\IZ[\La] \to \IZ\,,\quad &\tr\big\{ \sum_{i_1,\dots, i_\dd}x_1^{i_1}\dots x_{{\scaleto{\dd}{5pt}}}^{i_{\scaleto{\dd}{3.5pt}}}a_{i_1,\dots, i_{{\scaleto{\dd}{5pt}}}}\big\} =a_{0,\dots, 0}\,.
\end{align}
We use the same notation for the induced map \begin{align} \tr:S^{-1}R/R=\IQ[\La]/\IZ[\La]\longrightarrow\IQ/\IZ \end{align} defined by taking the coefficient of the identity monomial modulo $\IZ$.
\end{remark}
\vspace{3mm}

\begin{definition}\label{def_lagr}
    A \textit{sublagrangian} $L$ of a non-singular $\ep$-hermitian linking form $(P,\l)$ over $(R,S)$ is a submodule $i:L\hookrightarrow P$ such that 
\begin{itemize}
    \item $L$, $P/L$ are $(R,S)$-modules;
    \item  $L$ is isotropic, i.e.,
$i^{\vee}\circ\l\circ i=0$, or equivalently $\hl(L,L)=0$;
    \item The morphism of $R$-modules $i^{\vee}\circ \l :P\to L^{\vee}$ is surjective.
\end{itemize}
The annihilator of a sublagrangian $L$ in $(P,\l)$ is the submodule $L^{\perp}\coloneqq \ker  ( i^{\vee}\circ \l\,)$. A $\textit{lagrangian}$ is a sublagrangian $L$ such that $L^{\perp}=L$. In particular, a lagrangian $L$ fits into an exact sequence 
\begin{align}
\label{eq:lagr_ex_seq}
\begin{tikzcd}[ampersand replacement=\&, column sep=27pt]
	0 \arrow[r] \& L  \arrow[r, "i"] \& P \arrow[r, "i^{\vee}\circ \l"] \& L^{\vee} \arrow[r] \& 0\,.
\end{tikzcd}
\end{align}
\end{definition}
\begin{example}\label{example_3} 
Let $M$ and 
\begin{align}
    \hl = \begin{pmatrix} 0 & +\frac{1}{2}\\ -\frac{1}{2} & -\frac{x+x^{-1}}{4} \end{pmatrix}
\end{align} 
be the $(R,S)$-module and the form from Example \ref{example_2}. Then $L_1=Rm_1$ is a lagrangian. 
One can check that the only isotropic submodule different from $L_1$ is given by 

\begin{align} L_2=R(2m_2)=R((x+x^{-1})m_1) \end{align}
Notice however that $L_2$ not even a sublagrangian: one has \begin{align} L_2^{\perp}=L_1, \qquad M/L_2\cong R/(2,x+x^{-1})\oplus R/2R. \end{align} The first summand has projective dimension two because $(2,x+x^{-1})$ is a regular sequence in $R$. \end{example}
\vspace{3mm}

\begin{example}\label{example_4}
 Let us consider the linking form $M$ defined by the matrices
\begin{align}  \partial = \begin{pmatrix} p & 0 & 0 & f \\ 0 & p & -\overline{f} & 0 \\ 0 & 0 & p & 0 \\ 0 & 0 & 0 & p \end{pmatrix}, \quad \hl = \frac{1}{p^2} \begin{pmatrix} 0 & 0 & p & 0 \\ 0 & 0 & 0 & p \\ -p & 0 & 0 & f \\ 0 & -p & -\overline{f} & 0 \end{pmatrix} 
\end{align} 
where $ f \in \mathbb{Z}[\La]$. Let $m_1, m_2, m_3, m_4$ be generators of $M$. Then it is easy to see that $L_1 = \langle m_1, m_2 \rangle \cong (\mathbb{Z}/p[\Lambda])^2$ is a lagrangian. Moreover, we can obtain $\pd$ by applying the Horseshoe Lemma to the following exact sequence
\begin{align}
\begin{tikzcd}[ampersand replacement=\&, column sep=27pt]
	0 \arrow[r]\& L_1  \arrow[r] \& M \arrow[r] \& L_1^{\vee}\cong (\IZ/p[\La])^{\oplus 2}\arrow[r] \& 0\,.
    \end{tikzcd}    
\end{align}

On the other hand, let us consider the module $M'=\coker \pd'$ defined by the matrix
\begin{align} 
\partial' = \begin{pmatrix} p & f & 0 & 0 \\ 0 & p & 0 & 0 \\ 0 & 0 & p & -\overline{f} \\ 0 & 0 & 0 & p \end{pmatrix}. 
\end{align}
We observe that $M' \simeq N \oplus N^\vee$ where $N$ is the module defined by \eqref{(2.3)}. Let $m'_1,\ldots, m_4'$ be the generators of $M'$. Then the mapping
\begin{align}  m'_1 \mapsto m_1, \quad m'_2 \mapsto m_4, \quad m'_3 \mapsto m_2, \quad m'_4 \mapsto m_3 
\end{align}
defines an isomorphism $M' \cong M$. Under this isomorphism, the summand $N=\langle m'_1,m'_2\rangle$ maps to $L_2=\langle m_1,m_4\rangle\subset M$. Thus $L_2$ is also a lagrangian in $M$ with respect to $\hl$.
\end{example}

\begin{remark}
    We discuss the applications of $(R,S)$-modules to the generalized locally-flippable separators and generalizations of the stabilizer codes in \S \ref{sec:lattice_pauli_group}. As opposed to the free $\IZ/q[\La]$-modules having to do with the ordinary Pauli stabilizer codes \cite{haah2013}, there are many non-isomorphic $(R,S)$-modules with the same minimal numbers of generators. In particular, any two lagrangians with respect to a linking form need not be isomorphic as abstract modules. In Example \ref{example_4},  we observe, by setting $f(x) = x + x^{-1}$, that $L_2 \not\cong L_1 = (\mathbb{Z}/p[\Lambda])^2$ as abstract $R$-modules.
\end{remark}

\begin{remark}\label{remark:topological charges} 
Assume that $L\xhookrightarrow{i}P$ is isotropic and coisotropic in a non-singular $\ep$-hermitian linking form $(P,\l)$, so that $\Im i=\ker(i^{\vee}\circ\l)$. Then $L$ is a lagrangian if and only if both $L$ and $P/L$ are $(R,S)$-modules. Indeed, applying $\Hom_R(-,S^{-1}R/R)$ to \begin{align} 
\begin{tikzcd}[ampersand replacement=\&, column sep=27pt] 0 \arrow[r]\& L \arrow[r,"i"]\& P \arrow[r]\& P/L \arrow[r]\& 0
\end{tikzcd}
\end{align} gives 
\begin{align} P^{\vee}\longrightarrow L^{\vee}\longrightarrow \ext_R^1(P/L,S^{-1}R/R)\longrightarrow \ext_R^1(P,S^{-1}R/R)=0. 
\end{align} 
If $P/L$ is an $(R,S)$-module, Proposition~\ref{prop:top_charges} makes the middle obstruction vanish, so $i^{\vee}\circ\l:P\to L^{\vee}$ is surjective. Proposition~\ref{prop:top_charges} also identifies the condition that $L$ and $P/L$ are $(R,S)$-modules with the vanishing of their groups $\ext_R^j(-,S^{-1}R/R)$ for all $j\geqslant1$.
\end{remark}

\begin{definition}
A non-singular $\ep$-hermitian linking form $(M,\l)$ over $(R,S)$ is \textit{hyperbolic} if it admits a lagrangian and it is \textit{stably hyperbolic} if there exist hyperbolic forms $(M_1,\l_1)$ and $(M_2,\l_2)$ such that $(M\oplus M_1,\l\oplus \l_1)=(M_2,\l_2)$. 
    
\end{definition}
We define \begin{align}
    Q^{\ep}(R,S)=\{b\in S^{-1}R\,|\,b-\ep \bar b=a-\ep \bar a\,,a\in R \}/R\,.
\end{align}
An $\ep$-hermitian linking form over $(R,S)$ is \textit{even} if 
\begin{align}
    \hl(x,x)\in  Q^{\ep}(R,S)\quad \mbox{for all}\quad x\in P\,.
\end{align}

For example, any even anti-hermitian (anti-symmetric) linking form over $(\IZ,\IZ-\{0\})$ is isomorphic to $H^{-}(\CD)$ for some finite abelian group $\CD$, by the results of Wall \cite{WALL1963281}. We are aiming at applications to lattice systems with bosonic on-site degrees of freedom where even forms, like $\h^{-}(M)$, suffice. We discuss potential applications of odd forms to fermionic systems in Appendix \ref{app:majorana}.

\subsection{Linking formations} 
\begin{definition}\label{def:linking_formation}
    An (even) $\ep$-hermitian linking formation over $(R,S)$ is a tuple $(P,\l;L_1,L_2)$ consisting of an (even) non-singular $\ep$-hermitian linking form $(P,\l)$ over $(R,S)$, a lagrangian $L_1$ and a sublagrangian $L_2$ of $(P,\l)$. The linking formation is non-singular if $L_2$ is a lagrangian. An isomorphism of linking formations $(P,\l;L_1,L_2)$ and $(P',\l';L_1',L_2')$ is an isomorphism of linking forms $f:(P,\l)\xrightarrow{\sim}(P',\l')$ such that
\begin{align}
    f(L_1)=L_1'\,,\quad f(L_2)=L_2'\,.
\end{align}
\end{definition}

 We define the monoid $V^{ev}(R,S,\ep)$ of the isomorphism classes of \textbf{even} $\ep$-hermitian non-singular linking formations over $(R,S)$ with the following binary operation
    \begin{align}\label{eq:direct_sum_formation}
        [P,\l;L_1,L_2]+[ P',\l';L_1',L_2']=[ P\oplus P',\l\oplus \l';L_1\oplus L_1',L_2\oplus L'_2]\,.
    \end{align}

\begin{example} Consider $M$, $\l$, $L_1$, and $L_2$ from Example \ref{example_4}. This tuple is a non-singular formation. \end{example}

\begin{example}
Let $\a$ be an automorphism of a non-singular $\ep$-hermitian form $(P,\l)$ over $(R,S)$ and let $L\subset P$ be a lagrangian. Then, $(P,\l;L,\a(L))$ is a non-singular $\ep$-hermitian linking formation. In particular, if we take $(P,\l)=\hm((R/p)^q)$ and $L=(R/p)^q$ as the standard anti-hermitian form, $\a$ has the physical meaning of a Clifford QCA. 
\end{example}

If $K$ is a sublagrangian of $(P,\l)$, we write
\begin{align}\label{eq:sublagrangian_reduction_form}
    P_K\coloneqq K^{\perp}/K
\end{align}
for the reduced module and denote the induced linking form by $\l_K$; its pairing is
\begin{align}
    \hl_K([p],[q])\coloneqq\hl(p,q)\,.
\end{align}

\begin{definition}[\cite{Ranicki1981ExactSI}, p. 339]\label{def:witt_linking_formations}
    The \textit{Witt group} of even $\ep$-hermitian non-singular linking formations over $(R,S)$, denoted by $M^{ev}(R,S,\ep)$, is the quotient of $V^{ev}(R,S,\ep)$ by the following relations:
    \begin{itemize}
        \item $[ P,\l;L_1,L_2 ]+ [ P,\l;L_2,L_3 ]=[ P,\l;L_1,L_3 ]$
        \item For a formation $(P,\l;L_1,L_2)$, if there is a sublagrangian $K$ of $(P,\l)$ such that $K\subseteq L_2$ and $L_2/K$ is an $(R,S)$-module and such that $L_1\cap K=0 $, $P=K^{\perp}+L_1$, then 
\begin{align}\label{eq:stable_eq_of_formations}
    [P,\l;L_1,L_2]=[K^{\perp}/K,\l^{\perp}/\l;L_1\cap K^{\perp},L_2/K]\,.
\end{align}
\item For a formation $(P,\l;L_1,L_2)$, if there is a sublagrangian $K$ of $(P,\l)$ such that $K\subseteq L_1$ and $K\subseteq L_2$ with $L_1/K$ and $L_2/K$ being $(R,S)$-modules, then 
\begin{align}
    [P,\l;L_1,L_2]=[K^{\perp}/K,\l_K;L_1/K,L_2/K]\,.
\end{align}
    \end{itemize}
\end{definition}

The inverse of $[ P,\l;L_1,L_2]$ in $M^{ev}(R,S,\ep)$ is $[ P,\l;L_2,L_1]$ and all formations $( P,\l;L,L)$ are trivial in $M^{ev}(R,S,\ep)$.

\section{Twisted quasi-Pauli groups and local operator algebras}\label{sec:unitary_groups}

 The purpose of the section is to introduce groups of unitaries generating algebras of local operators from which we later build stabilizer Hamiltonians. 

\subsection{Pauli group} \label{Sec:Pauligroup}  
Let $\CD$ be a finite abelian group of order $q$. We use $\CD$ to label a basis $\{|g\ra\}_{g\in \CD}$ for
$\IC^q$ and define the ``shift" operators $\{X_g\}_{g\in \CD}$ acting via
\begin{align}\label{eq:GenX}
    X_h|g\ra=|h+g\ra\,.
\end{align}
Let $\CD^{\vee}=\Hom(\CD,\IQ/\IZ)$ be the Pontryagin dual of $\CD$ which is isomorphic to the group of characters of $\CD$, and to $\CD$ itself (non-canonically). For each $\psi \in \CD^{\vee}$, we define a ``clock" operator via
\begin{align}\label{eq:GenZ}
    Z_{\psi}|g\ra=e^{2\pi i \psi(g)}|g\ra\,.
\end{align}
The shift and clock operators satisfy the following commutation relations
\begin{align}\label{eq:PauliRels}
    [Z_{\psi},X_h]\coloneqq(Z_{\psi})^{-1} (X_h)^{-1} Z_{\psi} X_h=e^{2\pi i \psi(h)}\,.
\end{align}
We associate an ordered product of a shift and a clock $X_gZ_{\phi}$ with an element $(g,\phi)\in\CD\oplus\CD^{\vee}$. Recall the standard $\IQ/\IZ$-valued alternating form $\hl_0$ on $\CD\oplus\CD^{\vee}$: 
\begin{align}
\hl_0((g,\phi),(h,\psi)) = \phi(h)-\psi(g), \qquad g,h\in\CD, \quad \phi,\psi\in\CD^{\vee}. 
\end{align}
Then the commutator of two ordered products of shifts and clocks is \begin{align} 
[X_gZ_{\phi},X_hZ_{\psi}] = e^{2\pi i\hl_0((g,\phi),(h,\psi))}.
\end{align}
We define the \textit{Pauli group} as the group generated by the shifts, clocks, and phases
\begin{align}\label{def:PauliGroup}
    \textbf{P}(\hm(\CD);\CD)\coloneqq \{\omega X_{g}Z_{\psi}\,|\,\omega\in U(1),g\in \CD,\psi\in \CD^{\vee}\}\,.
\end{align}
Our notation $\textbf{P}(\hm(\CD);\CD)$ contains the information about the abstract group generated by the shifts and clocks and a choice of representation --- the Pauli operators act on the Hilbert space labeled by $\CD$. In this form and in similar context, the Pauli group appeared in \cite{moses2024cliffordcircuitsnoncyclicabelian}.

\subsection{Lattice Pauli group}\label{sec:lattice_pauli_group}
For $\dd>0$, the algebra of local operators on $\La$ contains a group of local unitaries generated by the Pauli operators acting on a single site. Let $j,k\in \La$ be the lattice sites, then we define the one-site shifts $X_g^j$ and clocks $Z_{\psi}^k$ \footnote{By $X^j_g$ we mean an operator of the form ``$\ldots\otimes  \id \otimes X_g\otimes \id \otimes \ldots$" which is identity everywhere except the site $j$.}:
\begin{align}\label{eq:commutator}
[Z_{\psi}^j,X_g^k]=\begin{cases}
     1\,,\hspace{36pt}\,\, j\neq k\,,\\
     e^{2\pi i \psi(g)}\,,\hspace{4mm} j=k\,.
 \end{cases}
\end{align}
The coordinate of the site $j\in \La$ is a multi-index $(j_1,\ldots ,j_{\dd})\in \IZ^{\dd}$. Then, an ordered product $X_g^kZ_{\psi}^j$ is associated with an element 
\begin{align}
    (g\, x_{1}^{k_1}\ldots x_{\dd}^{k_{\dd}},\psi \, x_{1}^{j_1}\ldots x_{\dd}^{j_{\dd}})\in \CD[x_1,x_1^{-1},\ldots ,x_{\dd},x_{\dd}^{-1} ]\oplus \CD^{\vee}[x_1,x_1^{-1},\ldots ,x_{\dd},x_{\dd}^{-1} ]\,.
\end{align}
A product of a finite number of the single-site clocks and shifts, considered up to a phase, is associated with the general element of $\CD[\La]\oplus \CD^{\vee}[\La]$. We define the \textit{lattice Pauli group} as a group generated by phases and single-site Pauli operators
\begin{align}
    \textbf{P}(\CD[\La])\coloneqq\la\omega, X_{g}^j,Z_{\psi}^k\,:\,\omega\in U(1),g\in \CD,\psi\in \CD^{\vee},j,k\in \La\ra\,.
\end{align}
  The commutation relations between arbitrary elements of $\textbf{P}(\CD[\La])$ can be expressed through the standard anti-hermitian linking form $\hm(\CD[\La])$. If $x,y\in \hm(\CD[\La])$ are the elements representing a pair of the Pauli operators (up to a phase);  the commutator of such operators is given by $e^{2\pi i \tr\{\hl_0(x,y)\}}$. In particular, we reproduce \eqref{eq:commutator} in this way.

\subsection{Local bases and $(R,S)$-modules}\label{ref:local_bases}

In the case of the Pauli groups, we associated each on-site qudit with the group ring $\IC[\CD]$ and we automatically obtained the action of $\CD$ by shifts. General $(R,S)$-modules instead give translation-covariant bases whose compactified dimensions agree with those of systems having finite-dimensional degrees of freedom per unit cell.

Given an $(R,S)$-module $L$, we introduce the Hilbert space of square-integrable sequences $\ell^2(L)$\footnote{The only difference being that we index our sequences by $L$ rather than $\mathbb N$.} and  equip it with the orthonormal basis $\{|m\ra\}_{m\in L}$ such that $\la m_1\,|\,m_2\ra=\delta(m_1-m_2)$, where $\delta$ is the indicator function. Any $(R,S)$-module $L$ is a countable abelian group and $\ell^2(L)$ is a separable Hilbert space (in this section it is especially important to remember that $R=\IZ[\La]$ and $S=\IZ-\{0\})$, infinite-dimensional if $\dd>0$. 

If we start with a translation-invariant system on $\La$, we can compactify it on a torus and impose periodic boundary conditions. We use $L_{\tau}$ to label the basis for the Hilbert space on a $\dd$-dimensional torus of length $\tau$. If $[L]=q\in K_0(T)$, then $\dim_{\IC}\ell^2(L_{\tau})=q^{\tau^{\dd}}$ by Proposition \ref{prop_counting}. Hence, non-canonically as Hilbert spaces, \begin{align} \ell^2(L_{\tau}) \cong \bigotimes_{x\in(\IZ/\tau)^{\dd}}\IC^q. \end{align} This dimension count does not by itself specify an onsite tensor-product realization of the represented operators.

\subsection{Quasi-Pauli groups and standard anti-hermitian linking forms over $(R,S)$}

Having defined the Hilbert space $\ell^2(L)$, we automatically obtain the shift operators defined above
and the clock operators labeled by $L^{\vee}$:
\begin{align}\label{eq:quasi_clock}
    Z_{w}|m\ra=e^{2\pi i \tr \{w(m)\}}|m\ra\,,\quad m\in L,w\in L^{\vee}\,.
\end{align} 
The shift and clock operators satisfy the commutation relations 
\begin{align}
    [Z_w,X_m]=e^{2\pi i \tr \{w(m)\}}\,.
\end{align}
The commutator of ordered products is expressed in terms of the standard anti-hermitian form $\hl_0^{-}$ on $L\oplus L^{\vee}$: 
\begin{align} 
[X_{m_1}Z_{w_1},X_{m_2}Z_{w_2}] = e^{2\pi i\tr\hl_0((m_1,w_1),(m_2,w_2))} = e^{2\pi i\tr\{w_1(m_2)-w_2(m_1)\}}, \quad m_1,m_2\in L,\; w_1,w_2\in L^{\vee}. 
\end{align}
 We will call the subgroup of the group of unitary automorphisms of  $\ell^2(L)$ generated by $Z_w,X_m$  \textit{the quasi-Pauli group} and denote by $\textbf{P}(\hm(L);L)$.

\subsection{Twisted quasi-Pauli groups}\label{sec:twisted_quasi_pauli}

The construction of Weyl operators  by a symplectic form over $\IR$ or $\IC$ and their Schrodinger representation by a choice of polarization is a classical subject, see e.g. \cite{Howe1980OnTR, Folland,denittis2025topologicaldescriptionpureinvariant}. 
In this section we define Weyl operators by any even non-singular anti-hermitian linking form $(P,\l)$ over $(R,S)$, a choice of a multiplier taking values in $\IQ/\IZ$, and a choice of lagrangian in $P$. Even though our construction is a logical generalization of the classical  Weyl operators, we are not aware of this generalization being mentioned in the existing literature.

  \subsubsection{Weyl system}
We wish to define a Weyl operator $W(p)$ for each $p\in P$ such that
\begin{align}\label{eq:Weyl_commutation}
    [W(p_1),W(p_2)]=e^{2\pi i \tr \{\hl(p_1,p_2)\}}\,,\quad p_1,p_2\in P\,.
\end{align}
Notice that the evenness of $\l$ guarantees the consistency of the relation \eqref{eq:Weyl_commutation}. Indeed, if $b\in S^{-1}R$ represents $\hl(p,p)$, evenness gives $b+\bar b=a+\bar a$ for some $a\in R$. Hence the constant coefficient of $b-a$ vanishes and 
\begin{align} 
\tr\{\hl(p,p)\}=0\in\IQ/\IZ.
\end{align} 

We define a (normalized) multiplier $\sigma$ as a function
   \begin{align}
       \sigma:P\times P\to \IQ/\IZ
   \end{align}
   such that 
   \begin{align}\label{bicharacter_relation}
       \sigma(p_1,p_2)-\sigma(p_2,p_1)=\tr\{\hl(p_1,p_2)\}\in \IQ/\IZ\,,\quad p_1,p_2\in P
   \end{align}
  and subject to the cocycle condition 
   \begin{align}
\sigma(p_1,p_2)+\sigma(p_1+p_2,p_3)=\sigma(p_2,p_3)+\sigma(p_1,p_2+p_3)
   \end{align}
for any triple $p_1,p_2,p_3\in P$, and subject to the normalization conditions
\begin{align}
    \sigma(0,p)=0\,,\;\sigma(p,0)=0\,,\;\sigma(p,-p)=0\,,\hspace{3mm} p\in P\,.
\end{align}
Note that, unlike $\tr\circ \hl$, we do not require the multiplier to be translation invariant. Given any such normalized multiplier $\sigma$, we can define the group of  Weyl operators $W(P,\lambda)$ by the following relations:
\begin{align}\label{eq:Weyl_relations}
    W(0)=1\,,\quad W(p_1)W(p_2)=e^{2\pi i \sigma(p_1,p_2)}W(p_1+p_2)\,,\quad W(p_1)^{-1}=W(-p_1)\,,\quad p_1,p_2\in P\,.
\end{align}

\subsubsection{Existence of the normalized multiplier.} Let us demonstrate that a normalized multiplier $\sigma$ exists for any $(P,\l)$. Essentially, this is the result of Exercise 8 in \S 3 Chapter IV of \cite{brown1982cohomology}. Here we consider an $(R,S)$-module $P$ as a $\IZ$-module, which is not finitely-generated unless $\dd=0$. The universal coefficients theorem states that there exists a short exact sequence
\begin{align}
    0\rightarrow\ext^1_{\IZ}(H_1(P,\IZ),\IQ/\IZ)\rightarrow H^2(P,\IQ/\IZ)\rightarrow \Hom_{\IZ}(H_2(P,\IZ),\IQ/\IZ)\rightarrow 0
\end{align}
We have $H_1(P,\IZ)\cong P$ since $P$ is already abelian and $\ext^1_{\IZ}(P,\IQ/\IZ)=0$ since $\IQ/\IZ$ is divisible. Further, $H_2(P,\IZ)\cong \Lambda^2P$ (see e.g. Theorem 6.4 (iii) in \cite{brown1982cohomology}) and the short exact sequence collapses to an isomorphism $H^2(P,\IQ/\IZ)\simeq \Hom_{\IZ}(\Lambda^2P,\IQ/\IZ)$ which acts on the cocycle by anti-symmetrization. In other words, there is an isomorphism between alternating forms and multipliers modulo coboundaries and any alternating form can be lifted to a multiplier. A standard computation shows that any cohomology class can be presented by a normalized multiplier. 

Assume that $P$ is such that $2\in \operatorname{Aut}(P)$. Then, we can choose an explicit multiplier which is also translation-invariant 
\begin{align}
    \sigma(x^j\cdot p_1,x^j\cdot p_2)=\sigma(p_1,p_2)
\end{align} where $j=(j_1,...,j_\dd)\in \Lambda$ and $x^j=x_1^{j_1}\cdots x_\dd^{j_\dd}$. For any non-singular anti-hermitian form $\l$ we derive the multiplier from another non-singular anti-hermitian form $\l\circ \frac{1}{2}$ such that $\sigma_{st}(p_1,p_2)=\tr\{\frac{1}{2}\hl(p_1,p_2)\}$. Since $\l\circ \frac{1}{2}$ is an anti-hermitian form, $\sigma_{st}$ is automatically a cocycle and it is translation-invariant. 

\subsubsection{Schrodinger representation}
Let $L$ be a lagrangian in $(P,\l)$ and $\sigma$ be a normalized multiplier. While $\hl|_{L}=0$, $\sigma$ need not vanish on $L$. Nevertheless, let us show that we can choose a normalized multiplier cohomologous to $\sigma$ which vanishes on $L$. We use the isomorphism $H^2(L,\IQ/\IZ)\cong \Hom_{\IZ}(\Lambda^2 L,\IQ/\IZ)$: since $\sigma(w_1,w_2)-\sigma(w_2,w_1)=0$ for all $w_1,w_2\in L$, $\sigma(w_1,w_2)=v(w_1+w_2)-v(w_1)-v(w_2)$ for some function $v:L\to \IQ/\IZ$. Since $\sigma$ is normalized, $v(0)=0$ and $v(-w)=-v(w)$. Then, we lift $v$ to a set-theoretic map $\tilde v:P\to \IQ/\IZ$ such that $\tilde v|L=v$ and $\tilde v(0)=0$ and $\tilde v(-p)=-\tilde v(p)$ for all $p\in P$. We define the new multiplier $\sigma(p_1,p_2)-\tilde v(p_1+p_2)+\tilde v(p_1)+\tilde v(p_2)$ which is normalized and which vanishes on $L$.

Let $L$ be a lagrangian in $(P,\l)$ and let $\sigma$ be a normalized multiplier vanishing on $L$. We define a (Schrodinger) representation of the Weyl operators on the Hilbert space
\begin{align}\label{eq:Schrodinger_rep}
    L^2_{(L,\sigma)}(P)
    \coloneqq
    \left\{
        f:P\to\IC
        \,\middle|\,
        f(x+w)=e^{-2\pi i\sigma(w,x)}f(x),
        \quad
        \sum_{[x]\in P/L}|f(x)|^2<\infty
    \right\}
\end{align}
with the following inner product 
\begin{align}
    (f,g)
    =
    \sum_{[x]\in P/L}
    \overline{f(x)}g(x).
\end{align}
Note that the terms in this sum do not depend on the choice of the representative of the class $x\in P/L$. Also note that the covariance condition in \eqref{eq:Schrodinger_rep} is well-defined as long as $\sigma|_L=0$.
The (unitary) action of the Weyl operators  on $L^2_{(L,\sigma)}(P)$ is given by the following rule
\begin{align}
    [W(p_1)f](x)=e^{2\pi i\sigma(x,p_1)}f(x+p_1)\,,\quad f\in L^2_{(L,\sigma)}(P)\,,\;x,p_1\in P\,.
\end{align}
In other words, we defined a representation induced from $L$ to $P$. The elements of the lagrangian act by diagonal unitaries: let $w\in L$, then 
\begin{align}
    [W(w)f](x)=e^{2\pi i\tr \{\hl(x,w)\}}f(x)\,.
\end{align}
Even though $L^2_{(L,\sigma)}(P)$ is a convenient representation, it is difficult to interpret it as a many-body Hilbert space. Let us demonstrate that there exists a unitary isomorphism $L^2_{(L,\sigma)}(P)\xrightarrow{\sim}\ell^2(L^{\vee})$. Let $\{|L\ra\}_{L\in L^{\vee}}$ be an orthonormal basis for $\ell^2(L^{\vee})$ and let $\pi:P\to L^{\vee}$ be the surjection, then we pick a section $s$ of $\pi$ and define a map
\begin{align}
U_s:L^2_{(L,\sigma)}(P)\xrightarrow{\sim}\ell^2(L^{\vee})\,,\quad U_s(f)=\sum_{m\in L^{\vee}}f(s(m))|m\ra\,.
\end{align}
Then, the inverse of $U_s$ is given by 
\begin{align}
U_s^{-1}|m\ra=\psi_m\,,\quad \psi_m(x)=e^{-2\pi i \sigma(x-s(m),s(m))}\delta(\pi(x)-m)\,,\quad m\in L^{\vee},x\in P\,.
\end{align}
Here $\{\psi_m\}_{m\in L^{\vee}}$ is an orthonormal basis for $L^2_{(L,\sigma)}(P)$  and $\delta(x)$ is the indicator function.
Notice that the cocycle condition on $\sigma$ as well $\sigma|_L=0$ gives  the following relation
\begin{align}\label{key_property_of_sigmasigma}
\sigma(w_1+w_2,x)=\sigma(w_1,x)+\sigma(w_2,x+w_1)
\end{align}
for any $w_1$, $w_2\in L$ and $x\in P$. This relation helps us to establish that $\psi_m\in L^2_{(L,\sigma)}(P)$ after we notice that $x-s(m)$ belongs to $L$ on the support of $\delta(\pi(x)-m)$.

If we choose another section $s'=s+v$ for $\pi$, where $v$ takes values in $L$, then the isomorphism 
\begin{align}
    U_{s+v}=D_{s,v}U_s\,,\quad D_{s,v}|m\ra=e^{-2\pi i\sigma(v(m),s(m))}|m\ra\,
\end{align}
shifts by a diagonal unitary $D_{s,v}$. Finally, we obtain the following action of the Weyl unitaries on $\ell^2(L^{\vee})$:
\begin{align}
    \widetilde{W}_s(p)|m\ra\coloneqq U_sW(p)U_s^{-1}|m\ra=e^{2\pi i \{\sigma(s(m-\pi(p)),p)-\sigma(s(m-\pi(p))+p-s(m),s(m))\}}|m-\pi(p)\ra\,.
\end{align}
Further changing $\sigma$ by a coboundary gives a unitarily equivalent representation.

\begin{definition}
    We define the \textit{twisted quasi-Pauli group} $\textbf{P}(P,\l;L)$ 
 as the unitary subgroup of $\End(\ell^2(L^{\vee}))$ generated by operators $\widetilde{W}_s(p)$ for all $p\in P$ and phases, for some choice of $s$.
 
\end{definition}
As discussed in \S\ref{ref:local_bases}, the $R$-module structure gives a basis-permuting translation action on $\ell^2(L^{\vee})$, and the compactified Hilbert spaces have the corresponding finite-dimensional counts. In the case when $(P,\l)=\hm(L)$ we can choose the section $s$ to be a homomorphism, and in that case we have a splitting of the Weyl operators into shifts and clocks and the twisted quasi-Pauli group reduces to the quasi-Pauli group. In general, $P$ is not a direct sum of the modules generated by the shifts and clocks respectively. The non-triviality of $P$ as an extension of $L^{\vee}$ by $L$ is implicitly encoded in the product of the Weyl operators \eqref{eq:Weyl_relations}. The group $\textbf{P}(P,\l;L)$ is the most general group of unitaries that we consider in this work. 

\begin{definition}[Algebra of local operators] \label{def:local_operator_algebra} Fix a lagrangian $L\subset(P,\l)$, a normalized multiplier $\sigma$ vanishing on $L$, and a section $s:L^{\vee}\to P$ as above. We define 
\begin{align} \CA(P,\l;L) \coloneqq \operatorname{span}_{\IC} \{\widetilde W_s(p)\mid p\in P\} \subseteq \End(\ell^2(L^{\vee})). 
\end{align} 
\end{definition}
Equivalently, $\CA(P,\l;L)$ is the unital $*$-algebra generated by the elements of the twisted quasi-Pauli group $\textbf{P}(P,\l;L)$.

\begin{remark} The elements of the algebra $  \CA(P,\l;L)$ are local in the following sense. Fix generators $e_1,\ldots,e_k$ of $P$. For a finite subset $\Gamma\subset\La$, let $\CA(P,\l;L)_\Gamma$ be the subalgebra generated by $\widetilde W_s(x^je_i)$ for $1\leqslant i\leqslant k$ and $j\in\Gamma$. Since $P$ is finitely generated and $S$-torsion, some nonzero integer annihilates $P$. Hence the subgroup generated by these finitely many labels is finite and $\CA(P,\l;L)_\Gamma$ is finite dimensional. The family is filtered under inclusion and 
\begin{align} 
\bigcup_{\Gamma\Subset\La}\CA(P,\l;L)_\Gamma = \CA(P,\l;L).
\end{align}
Relations among the generators can nevertheless make the same non-scalar operator belong to algebras associated with disjoint subsets. Thus this filtration does not determine an intrinsic minimal support for an operator. 
We think of the local algebra $\CA(P,\l;L)$ as obtained from the traditional algebra of local operators built from inclusions of local matrix algebras by imposing relations, see Remark \ref{phys_ex}.
\end{remark}

\subsection{Locally-flippable separators} The twisted quasi-Pauli groups are closely related to the locally-flippable separators from \cite{haah2023nontrivial}. A local separator is a set of mutually commuting local unitaries $\{\CZ_{i}\}_{i\in J}$ such that each element satisfies $(\CZ_{i})^q=1$ with some finite $q$ and such that this set of unitaries precisely separates the many-body Hilbert space. The separation condition means that if we diagonalize the elements of the separator and choose any set of the $q$-th roots of unity $\{e^{2\pi i n_{i}/q}\}_{i\in J}$, then the subspace $V$ of the many-body Hilbert space satisfying $\CZ_{i}V=e^{2\pi in_i/q}V$ is one-dimensional.  A local flipper $\{\CX_{i}\}_{i\in J}$ for the local separator 
$\{\CZ_{i}\}_{i\in J}$ is a set of local unitaries such that each element satisfies $(\CX_{i})^q=1$ and $[\CZ_i,\CX_j]=e^{2\pi i\delta_{i,j}/q}$. A local separator is called locally flippable (LFS) if a local flipper exists. Crucially, any LFS and the corresponding flipper together generate \textit{the full algebra of local operators $\CA$}.

For us, the index set $J$ for the LFS is the lattice $\La$ because we assume translation invariance. We obtain the LFS in the original form as in \cite{haah2023nontrivial} as a special case of our construction if $(P,\l)$ admits a lagrangian of the form  $L_0=\IZ/q[\La]$. Assume this is the case and let $e$ be a generator of $L_0$, then we obtain an LFS $\{\widetilde{W}_s(x^j\cdot e)\}_{j\in \La}$. Indeed, this set separates $\ell^2(L_0^{\vee})$. Let $\hat t$ be a generator for $P/L_0$ and let $t\in P$ be a representative of that class. Then, $[\widetilde{W}_s( e),\widetilde{W}_s(t)]=e^{2\pi i/q}$ and $\{\widetilde{W}_s(x^j\cdot t)\}_{j\in \La}$ is the flipper for the LFS above. 

Twisted quasi-Pauli groups provide a mild generalization of the LFS of \cite{haah2023nontrivial} by allowing more general $(R,S)$-modules than $\IZ/q[\La]$. Given a lagrangian $L$ in $(P,\l)$ and choosing a set of generators $\{e_{\a}\}_{\a\in I}$, we find a set of local mutually commuting unitaries $\{\widetilde{W}_s(x^j\cdot e_{\a})\}_{j\in \La, \a\in I}$, with each element having a finite order, and separating $\ell^2(L^{\vee})$. Indeed, there is a perfect pairing between $L$ and $L^{\vee}$ taking values in $\IQ/\IZ[\La]$. By choosing a generating set for $P/L$, we obtain a set of local unitaries (not necessarily commuting) that ``flip'' any single basis element of $\ell^2(L^{\vee})$. Let us call this generalization of LFS and local flippers \textit{quasi-LFS} and \textit{quasi-flippers}. 

The quasi-LFS together with their quasi-flippers generate the algebra
of local operators $\CA(P,\l;L)$ similarly to how the LFS together
with their local flippers generate the ordinary algebra of local observables of a spin system. One way of seeing this is by considering a compactification of $\La$ to a finite torus.  For each compactification on a torus of length $\tau$, we obtain exactly $q^{\tau^{\dd}}$ of mutually commuting operators, the ``clocks'', and $q^{\tau^{\dd}}$ of ``shift'' operators, using Proposition \ref{prop_counting}. The commutant of the algebra generated by such clocks and shifts is trivial. Indeed, since the clocks generate a maximally commutative subalgebra, the commutant must belong to that algebra. Then, commutation with all the shifts forces the commutant to be trivial. In the finite-dimensional case, we can use the double commutant theorem, which demonstrates that the shifts and clocks generate the full matrix algebra of a system compactified on a torus.

\section{Stabilizer codes}\label{ref:classification}
In this section we define the algebraic stabilizer codes classified in this paper, as well stabilizer groups, and corresponding local commuting-projector Hamiltonians,
and identify invertible codes with the linking formations of
Section~\ref{sec:linking forms}.

\subsection{Stabilizer codes}

The central definition of this work is of a stabilizer code. 
\begin{definition}[Stabilizer code]\label{def:stabilizer_code}
A \textit{stabilizer code} in $(P,\l)$ is a finitely generated
isotropic $R$-submodule $F\subset P$. A stabilizer code is called \textit{locally topologically ordered} if $F^{\perp}=F$.
\end{definition}
This definition is a generalization of the more conventional definition of a Pauli stabilizer code as a subset of a free module over $R/n$, see e.g. \cite{haah2013,haah2021classification}. We later show that every such code is realized in the ground space of at least one class of Hamiltonians, the class of frustration-free commuting-projector Hamiltonian on a local operator algebra in the sense of Definition \ref{def:local_operator_algebra}.

The quotient of $\textbf{P}(P,\l;L_1)$ by its central subgroup of phases is
canonically identified with the additive group of $P$. We denote the
corresponding phase-free label map by
\begin{align}
    \operatorname{lab}:\textbf{P}(P,\l;L_1)\longrightarrow P,
    \qquad
    \operatorname{lab}(\omega\widetilde W_s(p))=p\quad \mbox{for}\quad \omega\in U(1)\,.
\end{align}

\begin{definition}[Stabilizer group]
\label{def:stabilizer_group}
A \textit{stabilizer group} $\CS$ is a subgroup of
$\textbf{P}(P,\l;L_1)$ such that
\begin{align}\label{eq:no_phases}
    \CS\cap U(1)\id=\{\id\}
\end{align}
and such that $F_{\CS}\coloneqq\lab(\CS)\subset P$ is a stabilizer code. 
\end{definition}
A stabilizer group is thus a concrete unitary presentation of a stabilizer code. Condition \eqref{eq:no_phases} implies that
$\lab|_{\CS}:\CS\to F_{\CS}$ is an isomorphism; in particular, $\CS$ is
abelian. Note that the stabilizer group carries the action of $\La$.

\begin{definition}[Stabilizer presentation]
    Let $F\subset P$ be a stabilizer code. A \textit{stabilizer presentation} of $F$ in $\CA(P,\l;L_1)$ is a homomorphism 
\begin{align}\label{eq:splitting_over_code}
    \iota:F\longrightarrow
    \textbf{P}(P,\l;L_1)\subseteq\CA(P,\l;L_1)
\end{align}
from the additive group of $F$ to the multiplicative group of twisted quasi-Pauli operators which  satisfies $\operatorname{lab}\circ\iota=\id_{F}$.
\end{definition}
      Observe that $\iota$ is injective, $\iota(F)\cap U(1)\id =\{\id\}$. Therefore, $\iota(F)$ is a stabilizer group.

\begin{remark}
If $2\in\operatorname{Aut}(P)$, then we choose use the standard multiplier
\begin{align}
    \sigma_{st}(p,q)
    =
    \tr\left\{\frac{1}{2}\hl(p,q)\right\},
\end{align}
with $\sigma_{st}|_{F\times F}=0$ because $F$ is isotropic. In this case 
\begin{align}
    \iota(f)=\widetilde W_s(f)
\end{align}
is a stabilizer presentation.
\end{remark}

\begin{proposition}\label{prop:existence_stabilizer_presentation}
For any stabilizer code $F$, there exists a stabilizer group associated with
$F$. The stabilizer presentations of
$F$ are in one-to-one correspondence with functions
$v:F\to\IQ/\IZ$ satisfying
\begin{align}
    \sigma(f,g)=v(f+g)-v(f)-v(g),
    \qquad f,g\in F,
\end{align}
via
\begin{align}
    \iota_v(f)=e^{2\pi i v(f)}\widetilde W_s(f).
\end{align}
Equivalently, the set of stabilizer presentations of $F$ is a torsor for
$\Hom_{\IZ}(F,\IQ/\IZ)$.
\end{proposition}
\begin{proof}
 Let $\sigma$ be the
normalized multiplier used to define the Weyl operators. Under the isomorphism
\begin{align}
    H^2(F,\IQ/\IZ)
    \cong
    \Hom_{\IZ}(\Lambda^2F,\IQ/\IZ)
\end{align}
from Section~\ref{sec:twisted_quasi_pauli}, the class of the restriction of
$\sigma$ to $F$ is represented by its anti-symmetrization
\begin{align}
    (f,g)\longmapsto
    \sigma(f,g)-\sigma(g,f)
    =\tr\{\hl(f,g)\}.
\end{align}
Since $F$ is isotropic, this form vanishes. Hence the restriction of $\sigma$
to $F$ is a coboundary: there exists a function $v:F\to\IQ/\IZ$ such that
\begin{align}\label{eq:multiplier_trivialization_on_code}
    \sigma(f,g)=v(f+g)-v(f)-v(g),
    \qquad f,g\in F.
\end{align}
Since $\sigma$ is normalized, we have $v(0)=0$. Define
\begin{align}
    \iota_v(f)
    \coloneqq
    e^{2\pi i v(f)}\widetilde W_s(f),
    \qquad f\in F.
\end{align}
Then \eqref{eq:multiplier_trivialization_on_code} and the Weyl relations give
\begin{align}
    \iota_v(f)\iota_v(g)=\iota_v(f+g),
    \qquad
    \lab(\iota_v(f))=f.
\end{align}
Thus $\iota_v$ is a group-homomorphic splitting of the phase-free label map
over $F$, and its image is a stabilizer group associated with $F$.

Conversely, let $\CS$ be a stabilizer group associated with $F$ and let
\begin{align}
    \iota_{\CS}=(\lab|_{\CS})^{-1}.
\end{align}
Fix one solution $v$ of
\eqref{eq:multiplier_trivialization_on_code} and the corresponding
presentation $\iota_v$. There is a unique map $\delta:F\to U(1)$ such that
\begin{align}
    \iota_{\CS}(f)=\delta(f)\iota_v(f).
\end{align}
Since both $\iota_{\CS}$ and $\iota_v$ are homomorphisms, $\delta$ is a
homomorphism. Since $F$ is torsion, the values of $\delta$ are roots of
unity, so there is a unique
\begin{align}
    u\in\Hom_{\IZ}(F,\IQ/\IZ)
\end{align}
such that $\delta(f)=e^{2\pi i u(f)}$. Hence
$\iota_{\CS}=\iota_{v+u}$. This proves the claimed bijection.

\end{proof}

\subsection{Commuting-projector Hamiltonians}\label{sec:comm_proj_hams}
We now associate a commuting-projector stabilizer Hamiltonian with a
stabilizer presentation. Fix a stabilizer code $F$, a presentation
$\iota$, and $R$-generators $f_1,\ldots,f_s$ of $F$. Let
$n_k$ denote the additive order of $f_k$, which is finite because $F$ is
$S$-torsion, and define
 \begin{align}\label{eq:translated_stabilizers_commute}
    h_{j,k}\coloneqq\iota (x^jf_k),
    \qquad j\in\La,\quad 1\leqslant k\leqslant s,
\end{align}
and $h_k\coloneqq h_{0,k}$. Since multiplication by $x^j$ and $\iota$ are
injective, $h_{j,k}$ has order $n_k$. The operators $h_{j,k}$ commute and
generate $\iota(F)$. Define
\begin{align}\label{eq:stabilizer_projector}
    \Pi_{j,k}
    \coloneqq
    \frac{1}{n_k}\sum_{l=0}^{n_k-1}h_{j,k}^{l},
    \qquad
    Q_{j,k}\coloneqq\id-\Pi_{j,k}.
\end{align}
Then the projectors $\Pi_{j,k}$ are mutually commuting, and each
$\Pi_{j,k}$ is the spectral projector onto the $+1$ eigenspace of
$h_{j,k}$. The
\textit{commuting-projector stabilizer Hamiltonian} associated with $\iota$
and the chosen generators, which are suppressed from the notation, is the
formal sum
\begin{align}\label{eq:stabilizer_hamiltonian}
    \CH_{\iota}
    =\sum_{j\in\La}\sum_{k=1}^{s}Q_{j,k}.
\end{align}

\begin{lemma}\label{lem:linear_independence_quasi_pauli}
Let $u_1,\ldots,u_N\in\textbf{P}(P,\l;L_1)$ have pairwise distinct
phase-free labels
\begin{align*}
    p_a\coloneqq\lab(u_a)\in P,
    \qquad 1\leqslant a\leqslant N.
\end{align*}
Then, $u_1,\ldots,u_N$ are linearly independent in
$\CA(P,\l;L_1)$.
\end{lemma}

\begin{proof}
Write $u_a=\omega_a\widetilde W_s(p_a)$ with $\omega_a\in U(1)$. By
\eqref{eq:Weyl_commutation}, conjugation by a twisted quasi-Pauli operator
acts diagonally:
\begin{align*}
    \widetilde W_s(q)u_a\widetilde W_s(q)^{-1}
    =
    e^{2\pi i\tr\{\hl(q,p_a)\}}u_a,
    \qquad q\in P.
\end{align*}
These eigenvalues separate distinct labels. Indeed, suppose that
$p_a\neq p_b$ and put $r=p_a-p_b$. By the non-singularity and
anti-hermiticity of $\l$, there is $q_0\in P$ such that
$\hl(q_0,r)\neq0$. Write
\begin{align*}
    \hl(q_0,r)=\sum_{j\in\La}a_jx^j
    \in(\IQ/\IZ)[\La].
\end{align*}
Choose $j_0\in\La$ such that $a_{j_0}\neq0$ and set
$q=x^{j_0}q_0$. Since $\hl$ is conjugate linear in the first variable,
\begin{align*}
    \tr\{\hl(q,r)\}
    =
    \tr\{x^{-j_0}\hl(q_0,r)\}
    =
    a_{j_0}\neq0.
\end{align*}
Consequently,
\begin{align*}
    e^{2\pi i\tr\{\hl(q,p_a)\}}
    \neq
    e^{2\pi i\tr\{\hl(q,p_b)\}}.
\end{align*}

Assume now that there is a nontrivial linear relation with the smallest
possible number of nonzero terms,
\begin{align*}
    \sum_{a=1}^{N}c_au_a=0,
    \qquad c_a\neq0.
\end{align*}
Necessarily $N\geqslant2$. Choose $q\in P$ separating $p_1$ and $p_2$
as above. Conjugating the relation by $\widetilde W_s(q)$ and subtracting
$e^{2\pi i\tr\{\hl(q,p_1)\}}$ times the original relation gives
\begin{align*}
    \sum_{a=2}^{N}
    c_a\left(
        e^{2\pi i\tr\{\hl(q,p_a)\}}
        -
        e^{2\pi i\tr\{\hl(q,p_1)\}}
    \right)u_a=0.
\end{align*}
The coefficient of $u_2$ is nonzero, so this is a nontrivial relation
with fewer terms, a contradiction.
\end{proof}

\begin{proposition}\label{prop:derivation}
The formal sum in \eqref{eq:stabilizer_hamiltonian} defines a derivation
\begin{align}\label{eq:derivation}
    D_{\CH_{\iota}}(a)
    \coloneqq
    \sum_{j\in\La}\sum_{k=1}^{s}[Q_{j,k},a]
    \in\CA(P,\l;L_1),
    \qquad
    a\in\CA(P,\l;L_1).
\end{align}
\end{proposition}

\begin{proof}
Let us consider a summand $[Q_{j,k},a]$ in the formal sum in \eqref{eq:derivation}. It suffices to consider $a=\widetilde W_s(p)$ for some $p\in P$. For
$0\leqslant\ell<n_k$, the Weyl commutation relation gives
\begin{align}
    h_{j,k}^{\ell}\widetilde W_s(p)
    -
    \widetilde W_s(p)h_{j,k}^{\ell}
    =
    \left(
        e^{2\pi i\tr\{\hl(\ell x^jf_k,p)\}}-1
    \right)
    \widetilde W_s(p)h_{j,k}^{\ell}.
\end{align}
Since the form is conjugate-linear in the first variable,
\begin{align}
    \hl(\ell x^jf_k,p)
    =
    \ell x^{-j}\hl(f_k,p).
\end{align}
The element $\hl(f_k,p)\in S^{-1}R/R$ is represented by a Laurent
polynomial with finite support. Hence
\begin{align}
    \tr\{\ell x^{-j}\hl(f_k,p)\}=0
\end{align}
for all but finitely many $j\in\La$. Therefore
\begin{align}
    [Q_{j,k},\widetilde W_s(p)]=0
\end{align}
for all but finitely many $j$. By linearity, the same holds for every
$a\in\CA(P,\l;L_1)$, so the displayed sum is finite. The Leibniz rule
follows from the Leibniz rule for commutators.
\end{proof}

Thus the Hamiltonian defines an algebra derivation on the local algebra. Since the label map is defined only on twisted quasi-Pauli operators, we set
\begin{align}\label{eq:hamiltonian_centralizer_labels}
    K_{\CH_{\iota}}
    \coloneqq
    \lab\left(
        \ker D_{\CH_{\iota}}
        \cap \textbf{P}(P,\l;L_1)
    \right)
    \subset P.
\end{align}
This is a set of labels unaffected by derivation $D_{\CH_{\iota}}$.

\begin{proposition}\label{prop:recover_code_from_hamiltonian}
Let $\CH_{\iota}$ be the commuting-projector Hamiltonian constructed from a
stabilizer code $F\subset P$. Then
\begin{align}\label{eq:hamiltonian_centralizer_equals_orthogonal}
    K_{\CH_{\iota}}=F^{\perp}.
\end{align}
In particular, $K_{\CH_{\iota}}$ reproduces $F$ if $F$ is locally topologically ordered.
\end{proposition}

\begin{proof}
Let $u_p\in\textbf{P}(P,\l;L_1)$ have label $p\in P$, and define
\begin{align}
    \chi_p(g)
    \coloneqq
    e^{2\pi i\tr\{\hl(g,p)\}},
    \qquad g\in F.
\end{align}
Then
\begin{align}
    \iota(g)u_p=\chi_p(g)u_p\iota(g).
\end{align}
Writing
\begin{align}
    g_{j,k}\coloneqq x^jf_k,
\end{align}
the definition of $Q_{j,k}$ gives
\begin{align}\label{eq:commutator_projector_quasipauli}
    [Q_{j,k},u_p]
    =
    \frac{1}{n_k}
    \sum_{\ell=1}^{n_k-1}
    \bigl(1-\chi_p(\ell g_{j,k})\bigr)
    u_p\iota(\ell g_{j,k}).
\end{align}

For fixed $p$, the set
\begin{align}
    I_p
    \coloneqq
    \left\{
        (j,k)
        \,\middle|\,
        \chi_p(g_{j,k})\neq1
    \right\}
\end{align}
is finite by the same argument as in Proposition \ref{prop:derivation}.

Grouping the finitely many nonzero terms in
\eqref{eq:commutator_projector_quasipauli} according to their stabilizer
label gives
\begin{align}\label{eq:derivation_grouped_by_stabilizer_label}
    D_{\CH_{\iota}}(u_p)
    =
    \sum_{g\in F\setminus\{0\}}
    c_p(g)
    \bigl(1-\chi_p(g)\bigr)
    u_p\iota(g),
\end{align}
where only finitely many coefficients are nonzero and
\begin{align}
    c_p(g)
    \coloneqq
    \sum_{\substack{
        (j,k)\in I_p,\ 1\leqslant\ell<n_k\\
        \ell g_{j,k}=g
    }}
    \frac{1}{n_k}
    \geqslant0.
\end{align}

Suppose that $(j,k)\in I_p$ and set $g=g_{j,k}$. The term with
$\ell=1$ shows that
\begin{align}
    c_p(g)>0,
\end{align}
while
\begin{align}
    1-\chi_p(g)\neq0.
\end{align}
Moreover, as $g$ varies, the operators $u_p\iota(g)$ have the distinct
labels $p+g$ and are therefore linearly independent by Lemma \ref{lem:linear_independence_quasi_pauli}. Consequently,
\begin{align}
    D_{\CH_{\iota}}(u_p)=0
    \quad\Longleftrightarrow\quad
    \chi_p(x^jf_k)=1
\end{align}
for every $j\in\La$ and every $1\leqslant k\leqslant s$.

Equivalently,
\begin{align}
    \tr\{\hl(x^jf_k,p)\}=0
\end{align}
for every $j$ and $k$. Since the form is conjugate-linear in its first
variable,
\begin{align}
    \hl(x^jf_k,p)=x^{-j}\hl(f_k,p).
\end{align}
The traces of all monomial translates detect every coefficient of an element
of $S^{-1}R/R$. Therefore the preceding condition is equivalent to
\begin{align}
    \hl(f_k,p)=0
    \qquad
    \text{for every }k.
\end{align}
Since the elements $f_1,\ldots,f_s$ generate $F$ over $R$, this is
equivalent to
\begin{align}
    p\in F^{\perp}.
\end{align}
This proves
\eqref{eq:hamiltonian_centralizer_equals_orthogonal}. If $F$ is locally
topologically ordered, then $F^{\perp}=F$, and the final assertion follows.
\end{proof}

So far, we have demonstrated that for any abstract algebraic code $F$ satisfying the local topological order condition, there exists a commuting projector stabilizer Hamiltonian realizing the code as a set of constraints imposed on the ground state of that Hamiltonian. The commuting-projector Hamiltonian in \eqref{eq:stabilizer_hamiltonian} is not a unique realization of the code in the sense that many other Hamiltonians can define the same ground space. For simplicity, we work with \eqref{eq:stabilizer_hamiltonian}, while one could define a more general class of stabilizer Hamiltonians for the same stabilizer code.

In the following remark we explain why the set of constraints imposed on the ground space of the commuting-projector Hamiltonian always has a solution, i.e. that such Hamiltonians are frustration-free. 

\begin{remark}[Relation with the usual definition of frustration freeness]
\label{remark:usual_frustr_freeness}
For a commuting-projector Hamiltonian in a fixed unital
$*$-representation, frustration freeness means that all local terms can be
minimized simultaneously, equivalently that their common kernel is nonzero.
Commutativity alone does not guarantee this, since commuting projections may
have trivial common kernel.

Since no representation of the infinite system is fixed a priori, let
$\overline{\CA(P,\l;L_1)}
\subseteq\mathcal B(\ell^2(L_1^\vee))$ denote the operator-norm closure of
the local algebra. We say that $\CH_\iota$ admits a
\textit{frustration-free ground state} if there is a state $\omega$ on
$\overline{\CA(P,\l;L_1)}$ such that
\begin{align}\label{eq:frustration_free_state}
    \omega(Q_{j,k})=0,
    \qquad j\in\La,
    \quad 1\leqslant k\leqslant s.
\end{align}
For a finite subset $\Gamma\subset\La$, define
\begin{align}
    \CH_{\iota,\Gamma}\coloneqq
    \sum_{j\in\Gamma}\sum_{k=1}^{s}Q_{j,k},
 \quad 
    P_{\Gamma}
    \coloneqq
    \prod_{j\in\Gamma}\prod_{k=1}^{s}\Pi_{j,k}\,.
\end{align}
Since the $Q_{j,k}$ are commuting projections, for every unital
$*$-representation $\pi$ of $\overline{\CA(P,\l;L_1)}$,
\begin{align}\label{eq:local_ground_subspace}
    \ker\bigl(\pi(\CH_{\iota,\Gamma})\bigr)
    =
    \bigcap_{j\in\Gamma}\bigcap_{k=1}^{s}
    \ker\bigl(\pi(Q_{j,k})\bigr)
    =\operatorname{im}\bigl(\pi(P_{\Gamma})\bigr).
\end{align}
Thus the local ground subspace is precisely the subspace on which every
local term is minimized to zero. This subspace is nonzero in the Schrodinger representation on
$\ell^2(L_1^{\vee})$. Indeed, expanding $P_{\Gamma}$ using
\eqref{eq:stabilizer_projector} gives a finite linear combination of the
operators $\iota(g)$, $g\in F$, with nonnegative coefficients and a strictly
positive coefficient of $\id$. Since operators with distinct labels are
linearly independent, $P_{\Gamma}\neq0$. Choose a unit vector in
$\operatorname{im}(P_{\Gamma})$ and let $\omega_{\Gamma}$ be the corresponding
vector state on $\overline{\CA(P,\l;L_1)}$. Directing the finite subsets of $\La$ by
inclusion, weak-$*$ compactness of the state space gives an accumulation
point $\omega$ satisfying
\begin{align}
    \omega(Q_{j,k})=0,
    \qquad j\in\La,
    \quad 1\leqslant k\leqslant s.
\end{align}
If $(\pi_{\omega},\mathcal H_{\omega},\Omega_{\omega})$ is its GNS
representation, then
\begin{align}
    \|\pi_{\omega}(Q_{j,k})\Omega_{\omega}\|^2
    =\omega(Q_{j,k}^2)
    =\omega(Q_{j,k})
    =0.
\end{align}
Consequently,
\begin{align}
    \CG_{\iota,\omega}
    \coloneqq
    \bigcap_{j\in\La}\bigcap_{k=1}^{s}
    \ker\bigl(\pi_{\omega}(Q_{j,k})\bigr)
    \subseteq\mathcal H_{\omega}
\end{align}
is a nonzero closed subspace containing $\Omega_{\omega}$. We regard it as
the ground subspace of the formal Hamiltonian \eqref{eq:stabilizer_hamiltonian}.
\end{remark}

Proposition~\ref{prop:existence_stabilizer_presentation} and
Remark~\ref{remark:usual_frustr_freeness} show that every stabilizer code
admits a frustration-free commuting-projector realization. The
classification below retains only the module $F$ and is independent of the
choice of presentation and of the chosen $R$-generators.

\subsection{Invertible stabilizer codes as lagrangians}
\label{sec:codes/lagrangians}

Physically, the local topological order condition only guarantees that the ground state of the corresponding commuting-projector Hamiltonian is locally unique. However, there ground space might might globally non-unique due to the presence of non-local symmetries detected by the topological charges of \cite{bombin2014structure,haah2013,haah2021classification,ruba2024homological}. In our beyond-Pauli setting, the appropriate generalization is given by the Ext-groups discussed in Section~\ref{sec:top_charges}.

\begin{definition}[Topological-charge groups]
\label{def:stabilizer_topological_charges}
For an algebraic stabilizer code $F\subset P$, define
\begin{align}\label{eq:top_charges}
    \CQ_F^i&\coloneqq\ext_R^i(F,S^{-1}R/R),
    &
    \CQ_{P/F}^i&\coloneqq\ext_R^i(P/F,S^{-1}R/R),
    & i&\geqslant1.
\end{align}
\end{definition}

Since $F$ and $P$ are finitely generated and $P$ is $S$-torsion, both
$F$ and $P/F$ are finitely generated $S$-torsion modules. Proposition \ref{prop:top_charges} and Remark
\ref{remark:topological charges} give the following criterion.

\begin{proposition}\label{prop:invertibility_charges}
Let $F\subset P$ be an algebraic stabilizer code satisfying
$F^{\perp}=F$. Then the following conditions are equivalent:
\begin{enumerate}[label=(\roman*)]
    \item $F$ is a lagrangian in $(P,\l)$;
    \item $F$ and $P/F$ are $(R,S)$-modules;
    \item $\CQ_F^i=\CQ_{P/F}^i=0$ for every $i\geqslant1$.
\end{enumerate}
\end{proposition}

\begin{proof}
For finitely generated $S$-torsion modules, Proposition
\ref{prop:top_charges} identifies (ii) and (iii). Let
$i:F\hookrightarrow P$ be the inclusion. Under the hypothesis
$F=F^{\perp}$, Remark \ref{remark:topological charges} identifies (iii)
with exactness of
\begin{align}
    0\longrightarrow F\xrightarrow{\,i\,}P
    \xrightarrow{\,i^{\vee}\circ\l\,}F^{\vee}
    \longrightarrow0.
\end{align}
Together with isotropy, this is precisely the lagrangian condition in
Definition \ref{def_lagr}.
\end{proof}

\begin{definition}[Invertible stabilizer code]
\label{def:invertible_stabilizer_code}
Fix the reference lagrangian $L_1\subset P$. An \textit{invertible
stabilizer code} relative to $\textbf{P}(P,\l;L_1)$ is an algebraic
stabilizer code $L_2\subset P$ that is a lagrangian in $(P,\l)$.
Equivalently, by Proposition \ref{prop:invertibility_charges},
\begin{align}
    L_2=L_2^{\perp},
    \qquad
    \CQ_{L_2}^i=\CQ_{P/L_2}^i=0
    \quad\text{for every }i\geqslant1.
\end{align}
\end{definition}

An invertible stabilizer code is therefore specified by
\begin{align}\label{eq:code_formation}
    (P,\l;L_1,L_2),
\end{align}
where $L_1$ is the reference lagrangian and $L_2$ is the stabilizer
lagrangian. This is exactly an even non-singular anti-hermitian linking
formation over $(R,S)$. Conversely, the second lagrangian of any such
formation defines an invertible algebraic stabilizer code and, by Proposition
\ref{prop:existence_stabilizer_presentation}, admits a realization  by a commuting-projector
Hamiltonian. 

The diagonal formation $(P,\l;L,L)$ gives the direct analog of the product code in $\ell^2(L^{\vee})$. For the
standard hyperbolic form $\hm(L)=L\oplus L^{\vee}$, the formation
$(\hm(L);L,L^{\vee})$ gives its Fourier-dual code. We call them \textit{simple
codes}.

\subsection{Stabilization and condensation}
\label{sec:witt_equivalence_codes}

An invertible stabilizer code relative to
$\textbf{P}(P,\l;L_1)$ determines the datum
$(P,\l;L_1,L_2)$, where $L_1$ is the reference lagrangian and $L_2$ is
the stabilizer lagrangian. For concreteness, we realize the invertible codes by commuting-projector stabilizer Hamiltonians, as discussed in Section \ref{sec:comm_proj_hams}, while noting that more general stabilizer Hamiltonians exist.

Given commuting-projector stabilizer Hamiltonians $\CH_1$ and $\CH_2$, their stacking is given by
\begin{align}\label{eq:stacked_Ham}
    \CH_{12}=\CH_1\otimes\id+\id\otimes\CH_2.
\end{align}
On phase-free labels, stacking sends
\begin{align}
    (P,\l;L_1,L_2),\ (P',\l';L_1',L_2')
    \longmapsto
    (P\oplus P',\l\oplus\l';
      L_1\oplus L_1',L_2\oplus L_2').
\end{align}
It therefore agrees with direct sum of linking formations.

Consider a stack of invertible codes $(P,\l;L_1,L_2)$ and
$(P,\l;L_2,L_3)$. We use the same generators of the two copies of $P$
and compatible choices of Weyl representatives. Thus, for every
$p\in P$, the operators
\begin{align}
    \widetilde W^{(1)}(p)\in\textbf{P}(P,\l;L_1),
    \qquad
    \widetilde W^{(2)}(p)\in\textbf{P}(P,\l;L_2)
\end{align}
are two representations of the same abstract Weyl operator $W(p)$.
The layer-exchange automorphism is determined by
\begin{align}
    \gamma_{\mathrm{SWAP}}
    \bigl(\widetilde W^{(1)}(p)\otimes\id\bigr)
    &=
    \id\otimes\widetilde W^{(2)}(p),\\
    \gamma_{\mathrm{SWAP}}
    \bigl(\id\otimes\widetilde W^{(2)}(p)\bigr)
    &=
    \widetilde W^{(1)}(p)\otimes\id.
\end{align}
Broadly speaking, the layer-exchange automorphism $\gamma_{\mathrm{SWAP}}$ does not change the entanglement pattern of a stack of codes. More precisely, this is a simple example of a finite-depth quantum circuit, which, in particular, preserves the twisted quasi-Pauli groups. Therefore, we consider the codes related by $\gamma_{\mathrm{SWAP}}$ as equivalent.

The automorphism $\gamma_{\mathrm{SWAP}}$ sends the stabilizers labeled by $L_2$ in the first factor to the
stabilizers labeled by $L_2$ in the second factor, and the stabilizers
labeled by $L_3$ in the second factor to those labeled by $L_3$ in the
first. Hence, after the SWAP, the stacked code is
\begin{align}
    (P,\l;L_1,L_3)+(P,\l;L_2,L_2).
\end{align}
The second summand is the diagonal simple code and is trivial under
stabilization. Therefore,
\begin{align}\label{eq:Maslov_rel_codes}
    [P,\l;L_1,L_2]+[P,\l;L_2,L_3]
    =
    [P,\l;L_1,L_3].
\end{align}

We next explain the two sublagrangian reductions in Definition
\ref{def:witt_linking_formations}. These are the algebraic operations that we
interpret as condensation. Let $K\subset P$ be a sublagrangian and choose a
stabilizer presentation of $K$. Physically, we add a large penalty enforcing
the corresponding commuting constraints and restrict to their common $+1$
eigenspace. At the level of phase-free labels, an operator preserves this
subspace exactly when its label lies in $K^{\perp}$. Inside the constrained
subspace, labels differing by an element of $K$ induce the same effective
operator up to a phase. Consequently, the effective label module and its
commutator pairing are
\begin{align}\label{eq:physical_sublagrangian_reduction}
    P_K=K^{\perp}/K,
    \qquad
    \hl_K([p],[q])=\hl(p,q).
\end{align}
The additional hypotheses in the two reductions below ensure that the
reduced reference and stabilizer modules are again lagrangians in the category
of $(R,S)$-modules. A quotient by an arbitrary isotropic submodule need not
define another invertible stabilizer code.

First, suppose that the constraints are contained in the stabilizer
lagrangian, $K\subseteq L_2$, and that
\begin{align}\label{eq:first_condensation_hypotheses}
    L_2/K\text{ is an }(R,S)\text{-module},
    \qquad
    L_1\cap K=0,
    \qquad
    P=L_1+K^{\perp}.
\end{align}
Let $q_K:K^{\perp}\to P_K$ be the quotient map and set
\begin{align}
    \overline{L}_1
    \coloneqq q_K(L_1\cap K^{\perp})
    =\bigl((L_1\cap K^{\perp})+K\bigr)/K.
\end{align}
Because $K$ need not be contained in $L_1$, the reduced reference lagrangian
is $\overline{L}_1$, rather than $L_1/K$. The effective code datum is
\begin{align}\label{eq:condensation_reduction_code}
    [P,\l;L_1,L_2]
    =
    [P_K,\l_K;\overline{L}_1,L_2/K].
\end{align}
This is the first sublagrangian relation in Definition
\ref{def:witt_linking_formations}. Physically, \eqref{eq:condensation_reduction_code} says that one
may first impose a compatible collection of stabilizer constraints and then
describe the remaining code using only the operators that preserve the
constrained subspace.

Second, suppose that the same sublagrangian is contained in both
lagrangians,
\begin{align}
    K\subseteq L_1\cap L_2,
\end{align}
and that $L_1/K$ and $L_2/K$ are $(R,S)$-modules. The constrained part is
then common to the reference and stabilizer data and can be removed from
both. This gives
\begin{align}\label{eq:common_condensation_reduction_code}
    [P,\l;L_1,L_2]
    =
    [P_K,\l_K;L_1/K,L_2/K].
\end{align}
This is the common-sublagrangian relation in the Witt group. Unlike
\eqref{eq:condensation_reduction_code}, it is not part of Ranicki's stable
equivalence; it is an additional Witt relation. Notice in particular that no
condition of the form $P=K^{\perp}+L_i$ is imposed in
\eqref{eq:common_condensation_reduction_code}.

The simple codes introduced above are trivial under these relations. Taking
$K=L$ in \eqref{eq:common_condensation_reduction_code} gives
\begin{align}\label{eq:diagonal_simple_code_trivial}
    [P,\l;L,L]=0.
\end{align}
Combining this with \eqref{eq:Maslov_rel_codes} gives
\begin{align}\label{eq:inverse_code_formation}
    -[P,\l;L_1,L_2]
    =
    [P,\l;L_2,L_1].
\end{align}
For the standard hyperbolic form, taking $K=L^{\vee}$ in
\eqref{eq:condensation_reduction_code} gives
\begin{align}\label{eq:fourier_simple_code_trivial}
    [\hm(L);L,L^\vee]=0.
\end{align}

We call two finite stacks of invertible stabilizer code data \textit{Witt
equivalent} if one can pass from one to the other by a finite sequence of
isomorphisms and applications, in either direction and inside an arbitrary
stack, of the composition relation \eqref{eq:Maslov_rel_codes} and the two
reductions \eqref{eq:condensation_reduction_code} and
\eqref{eq:common_condensation_reduction_code}. We conclude that the invertible stabilizer codes up to the Witt equivalences are given by 
\begin{align}
    M^{ev}(R,S,-).
\end{align}

 The group $M^{ev}(R,S,-)$ also contains classes induced from
lower-dimensional systems. Section~\ref{sec:L-groups} removes these
contributions and computes the reduced groups $\operatorname{ISC}_{\dd}$.
\section{\textit{L} - groups}\label{sec:L-groups}

This section is devoted to the calculation of $M^{ev}(R,S,-)$. We briefly review the parts of the algebraic $L$-theory that are relevant to the computation of these Witt groups.  We recommend \cite{Ranicki1981ExactSI,ranicki1992algebraic} for the classical textbook account of $L$-theory and \cite{calmes2020hermitian1,calmès2025hermitian2,calmes2020hermitian3} for a modern approach. For a commutative ring with involution $R$, and for each $n\geqslant 0$, the symmetric (quadratic) $L$-group $\L^{s}_n(R,\ep)$ ($
\L^{q}_n(R,\ep)$) is the abelian  group of cobordism classes of $n$-dimensional $\ep$-hermitian ($\ep$-quadratic) Poincare complexes of FGP $R$-modules, see $\S1$ of \cite{Ranicki1981ExactSI}. Quadratic $L$-groups satisfy the $4$-fold periodicity (Proposition 1.2.3 in \cite{Ranicki1981ExactSI})
\begin{align}
   \L^{q}_n(R,\ep)\cong \L_{n+4}^q(R,\ep)\cong \L_{n+2}^q(R,-\ep)\,,\quad n\geqslant0\,.
\end{align}
This motivates the extension of the quadratic $L$-groups to the negative indices by the $4$-periodicity as was done in \cite{RanickiAlgThofSurgery1}. The symmetric $L$-groups are in general not periodic. Recent works \cite{calmes2020hermitian1,calmès2025hermitian2,calmes2020hermitian3} have given a definition to the genuine $L$-theory groups for all $n\in \IZ$, which we are going to use. In the positive degrees, they reproduce the groups of cobordism classes of hermitian complexes; in degrees $-1$ and $-2$ they are the even symmetric groups, while in degrees $n\leqslant-3$ they coincide with the quadratic $L$-groups. We will denote the genuine $L$-groups the same as the symmetric ones: 
\begin{align}\label{eq:SymLTh}
    \L^s_n(R,\ep)=\begin{cases}
       \L^{s,even}_{n+2}(R,-\ep)\,,\quad n=-2,-1\,,\\
        \L^q_{n}(R,\ep)\,,\quad \;\, n\leqslant -3\,.
    \end{cases}
\end{align}
where $\L^{s,even}$ are the even $L$-groups \cite{RanickiAlgThofSurgery1}.
 Crucially, $\L_0^s(R,\ep)$ and $\L^s_1(R,\ep)$ are isomorphic to the Witt group of $\ep$-hermitian forms and formations respectively, see Proposition 5.1 and 5.2 of \cite{RanickiAlgThofSurgery1}.  

Similar to a general cohomology theory, associating a relative cohomology group to a morphism, there are relative $L$-theory groups associated to a morphism of rings. For localizations $R\to S^{-1}R$, relative $L$-groups $\L_n^s(R,S,\ep)$ can be defined as abelian groups of cobordism classes of $n$-dimensional Poincare complexes that become null-homotopic upon localization \cite{Ranicki1981ExactSI}. In particular, $\L^{s}_1(R,S,\ep)$ is isomorphic to the group of even $\ep$-hermitian linking formations over $(R,S)$ (see p. 339 of \cite{Ranicki1981ExactSI}):
\begin{align}
    \L_1^s(R,S,\ep)=M^{ev}(R,S,\ep)\,.
\end{align}

Crucially, they fit into the localization exact sequence (Proposition 3.6.1 of \cite{Ranicki1981ExactSI}):
\begin{align}\label{eq:LocalizationLES}
    \ldots\lr \L^{s}_n(R,\ep)\lr {}_{\tilde{S}}\L^s_n(S^{-1}R,\ep)\lr \L_n^s(R,S,\ep)\lr \L^s_{n-1}(R,\ep)\lr\ldots
\end{align}
Here, ${}_{\tilde S}\L^s_n(S^{-1}R,\ep)$ denotes the $L$-group with a fixed $K_0$-theory class: it is defined similarly to $\L^s_n(S^{-1}R,\ep)$ except that all the FGP $S^{-1}R$-modules involved must come from FGP $R$-modules. In our applications, all FGP $S^{-1}R$-modules can be obtained from FGP $R$-modules by localization, $K_0(\IZ[\La])\cong K_0(\IQ[\La])\cong \IZ$. In other words, for our rings, ${}_{\tilde S}\L^s_n(S^{-1}R,\ep)=\L^s_n(S^{-1}R,\ep)$ so that we can drop the subscript.

For special rings, the $L$-groups play nicely with the 
Laurent extensions. Suppose that $R$ is a commutative ring with involution and $R[x,x^{-1}]$ is its Laurent extension such that $\bar x=x^{-1}$. Also suppose that $K_0(R)\cong K_0(R[x,x^{-1}])\cong  \IZ$. Then, there is an isomorphism (Example 1.3.11 \cite{calmes2020hermitian3}):
\begin{align}\label{eq:LsSplitting}
    \L^{s}_n(R[x,x^{-1}],\ep)= \L^{s}_n(R,\ep)\oplus  \L^{s}_{n-1}(R,\ep) .
\end{align}
 See also \cite{Ranicki1973II} for a proof of this relation in the quadratic range.

The long exact sequence \eqref{eq:LocalizationLES} for
$\L_1^s(\IZ[\La],\IZ\!-\!\{0\},-)$ splits into direct sums according
to the splitting in \eqref{eq:LsSplitting}. Let
\begin{align}
    G_\dd
    \coloneqq
    \L_1^s(\IZ[\IZ^\dd],\IZ\!-\!\{0\},-).
\end{align}
For $1\leqslant i\leqslant\dd$, let
\[
    \iota_i:
    \IZ[\IZ^{\dd-1}]
    \hookrightarrow
    \IZ[\IZ^\dd]
\]
be the coordinate inclusion whose image omits the Laurent variable
$x_i$. The image of $(\iota_i)_*$ consists of phases obtained from a
lower-dimensional lattice system. We define
\begin{align}\label{eq:reduced_phases}
    \operatorname{ISC}_\dd
    \coloneqq
    \begin{cases}
        0,
        & \dd\leqslant-1,\\[1mm]
        G_0,
        & \dd=0,\\[1mm]
        \coker\left(
        \displaystyle\bigoplus_{i=1}^{\dd}G_{\dd-1}
        \xrightarrow{\,\sum_i(\iota_i)_*\,}
        G_\dd
        \right),
        & \dd\geqslant1.
    \end{cases}
\end{align}
By naturality and iteration of \eqref{eq:LsSplitting},
\begin{align}
    G_\dd
    \cong
    \bigoplus_{I\subseteq\{1,\ldots,\dd\}}
    \L_{1-|I|}^s(\IZ,\IZ\!-\!\{0\},-).
\end{align}
The sum of the coordinate images is the sum of the terms indexed by
proper subsets $I$. Therefore,
\begin{align}
    \operatorname{ISC}_\dd
    \cong
    \L_{1-\dd}^s(\IZ,\IZ\!-\!\{0\},-).
\end{align}
Hence the localization exact sequence gives
\begin{align}\label{eq:sequenceCE}
    \ldots\lr \L^{s}_{1-\dd}(\IZ,-)\lr \L^s_{1-\dd}(\IQ,-)\lr \operatorname{ISC}_{\dd}\lr \L^s_{-\dd}(\IZ,-)\lr\ldots
\end{align}

The computation boils down to the $L$-groups of Dedekind domains with trivial involution, which are well-known, together with the morphisms between them \cite{wall1999surgery,RanickiAlgThofSurgery1,Ranicki1981ExactSI}. By Proposition 6.1 of \cite{RanickiAlgThofSurgery1}, 
\begin{align}
\L_n^s(\IZ,+)=\L^s_{n+2}(\IZ,-)\,,\quad \L_n^s(\IQ,+)=\L^s_{n+2}(\IQ,-)\,,\quad n\in \IZ\,,
\end{align}
so that we can shift the sequence and use only the $L$-groups with $\ep=+$. We define
\begin{align}
\L_n^s(R)\coloneqq  \L^s_n(R,+)\,,\quad \L_n^q(R)\coloneqq  \L^q_n(R,+)\,.
\end{align}
The Witt groups of rational bilinear forms are calculated in, e.g., \cite{milnor2013symmetric}:
\begin{align}
\L^q_n(\IQ)=\L_n^s(\IQ)=\begin{cases}
\IZ \oplus \bigoplus_{p\,\scaleto{\mbox{prime}}{7pt}}\CW^{sym}_p \,,\hspace{27pt}n\equiv 0\;\mbox{mod}\;4\,,\\
0\,,\quad\quad\quad\quad\quad\quad\quad\quad\quad \;\mbox{otherwise}\,.
\end{cases}
\end{align}

Here, the first summand corresponds to the signature of a symmetric form and  the subsequent summands are the Witt groups of symmetric forms over $\IF_p$:
\begin{align}
   \CW^{sym}_p= \begin{cases}
         \IZ/2\,,\quad\quad\quad\;\;\; p=2\,,\\
        \IZ/2\oplus \IZ/2\,,\quad p\equiv 1\;\mbox{mod}\;4\,,\\
        \IZ/4\,,\quad\quad \quad\quad p\equiv 3\;\mbox{mod}\;4\,.
    \end{cases}
\end{align}
 The remaining $L$-groups are calculated in Section 7 of \cite{RanickiAlgThofSurgery1}:
\begin{align}
\mbox{For}\;n\in\IZ\,,\quad \L^q_n(\IZ)=&\begin{cases}
    \IZ\,,\quad\quad\quad\quad\; n\equiv 0\;\mbox{mod}\;4\,,\\
    0\,,\quad\quad\quad\quad\;\, n\equiv 1\;\mbox{mod}\;4\,,\\
    \IZ/2\,,\quad\quad\quad\; n\equiv 2\;\mbox{mod}\;4\,,\\
    0\,,\quad\quad\quad\quad\;\; n\equiv 3\;\mbox{mod}\;4\,.
\end{cases}\\
\mbox{For}\;n\geqslant0\,,\quad \L^s_n(\IZ)=& \begin{cases}
    \IZ\,,\quad\quad\quad\quad\;\, n\equiv 0\;\mbox{mod}\;4\,,\\
    \IZ/2\,,\quad\quad\quad\;\, n\equiv 1\;\mbox{mod}\;4\,,\\
0\,,\quad\quad\quad\quad\;\,\,n\equiv 2\;\mbox{mod}\;4\,,\\
    0\,,\quad\quad\quad\quad\;\,\, n\equiv 3\;\mbox{mod}\;4\,.
\end{cases}\\
\mbox{For}\;n<0\,,\quad \L^s_n(\IZ)=&\begin{cases}
    0\,,\quad\quad\quad\quad\;\;\, n=-1,-2\,,\\
    \L_n^q(\IZ)\,,\quad\quad\;\; n\leqslant -3\,.
\end{cases}
\end{align}

Under the identification $\L^q_0(\IZ)\cong\IZ$ given by the signature divided by $8$, the map $\L^q_0(\IZ)\to \L_0^q(\IQ)$ is multiplication by $8$ on the signature summand $\IZ$. We obtain the following result
\begin{align}\label{eq:CliffordPhases}
\operatorname{ISC}_\dd=\begin{cases}\CW^{pt}\,,\;\;\,\, \dd\equiv 3\;\,\mbox{mod}\;4\; \&\; \dd>0\,,\\
  \IZ/2\,,\quad \dd\equiv 0\;\,\mbox{mod}\;4\; \&\; \dd>0\,,\\
  0\,,\quad\quad\; \mbox{otherwise}\,.\\
  \end{cases}
\end{align}
Here, $\CW^{pt}$ is the Witt group of quadratic linking forms on finite abelian groups
\begin{align}\label{eq:witt_groups_quad_forms}
\CW^{pt}= \bigoplus_{p\,\scaleto{\mbox{prime}}{7pt}}\CW^{pt}_p\,,\quad\quad
   \CW^{pt}_p= \begin{cases}
        \IZ/8\oplus \IZ/2\,,\quad p=2\,,\\
        \IZ/2\oplus \IZ/2\,,\quad p\equiv 1\;\mbox{mod}\;4\,,\\
        \IZ/4\,,\quad\quad \quad\quad p\equiv 3\;\mbox{mod}\;4\,.
    \end{cases}
\end{align}

\section{Spectrum of invertible stabilizer codes}\label{sec:Lspectra}
 The goal of this section is to introduce the notation and prove Proposition \ref{prob:GEM}. The reader familiar with spectra can skip the preliminaries and proceed to the end of the section. 

An $L$-theory is a functor from the category of rings with involution to the category of spectra \cite{ranicki1992algebraic,calmes2020hermitian1} such that the $L$-groups of a ring are homotopy groups of the associated spectrum. Ultimately, the sequence in \eqref{eq:sequenceCE} calculates the homotopy groups of the spectrum which is a cofiber of maps between two spectra. Below, we explicitly write down those spectra.

The category of spectra is a symmetric monoidal $(\infty,1)$-category when equipped with the smash product with the sphere spectrum as a unit. There are many versions of $L$-theories \cite{RanickiAlgThofSurgery1,ranicki1992algebraic} and we are going to use the modern version of the $L$-theory spectrum, described in \cite{calmes2020hermitian3}, which goes by the name ``genuine $L$-theory". We denote by  $\L^s(R,\ep)$ and $\L^{q}(R,\ep)$ the genuine symmetric and quadratic $L$-theories respectively. The non-negative homotopy groups of $\L^s(R,\ep)$ are isomorphic to the cobordism groups of $\ep$-symmetric Poincare complexes of $R$-modules while the negative groups satisfy \eqref{eq:SymLTh}, so that we have 
\begin{align}
    \L^s_n(R,\ep)=\pi_n\L^s(R,\ep)\,,\quad \L^q_n(R,\ep)=\pi_n\L^q(R,\ep)\,.
\end{align}

We denote $\L^s(R)\coloneqq \L^s(R,+)$ and $\L^q(R)\coloneqq \L^q(R,+)$. We recognize that the simplified version \eqref{eq:sequenceCE} of the localization long exact sequence \eqref{eq:LocalizationLES} is just a sequence associated to a map of spectra $\L^s(\IZ)\to \L^s(\IQ)$. Let $\operatorname{cofib}(\L^s(\IZ)\rightarrow \L^s(\IQ))$ be the cofiber, then the relative $L$-groups are given by
\begin{align}
    \L^s_n(\IZ,\IZ-\{0\},-)=\L^s_{n-2}(\IZ,\IZ-\{0\},+)=\pi_{n-2}\operatorname{cofib}(\L^s(\IZ)\rightarrow \L^s(\IQ))\,.
\end{align}

In principle, there can be many spectra with the same homotopy groups. However, in our case we can pin down the spectrum by its homotopy groups, and this turns out to be the simplest spectrum with prescribed homotopy groups -- a generalized Eilenberg-MacLane spectrum. Namely, given an abelian group $A$,
the Eilenberg-MacLane spectrum $HA$ is the unique spectrum which has only one non-zero homotopy group $\pi_0(HA)=A$. We will say that a spectrum $E$ is a generalized Eilenberg-MacLane spectrum if it can be represented as 
$$E=\bigvee_{i\in \mathbb Z}\Sigma^i HA_i$$
where $A_i$ are abelian groups. Note that any $H\bZ$ -- module is a generalized Eilenberg-MacLane spectrum, see, e.g. Proposition 5.3 in \cite{Casacuberta}.

\begin{remark}
Let  $p$ be a prime number, then the localization of $\IZ$ at $p$ is $ \IZ_{(p)}\coloneqq\{\frac{a}{b}\in \IQ\,|\,p\nmid b\}$. Let $Sp_{(p)}$ be a subcategory of the $p$-local spectra, the full subcategory of $Sp$ containing all spectra with homotopy groups which do not change after $\otimes_{\IZ}\IZ_{(p)}$. The embedding functor admits a left-adjoint
$$Sp\to Sp_{(p)}; \,\,\, E\mapsto E_{(p)}$$
which is called $p$-localization. More precisely, this functor can be described as a smashing  localization which maps $E$ to $M\IZ_{(p)}\wedge E$ (Proposition 2.4 of \cite{BOUSFIELD1979257}). Here $M\IZ_{(p)}$ is a Moore spectrum of $\bZ_{(p)}$. This implies two facts which will be important for us:
\begin{itemize}
    \item If the homotopy groups of a spectrum are torsion groups and contain only $p^k$-torsions then the spectrum is automatically $p$-local;
    \item  Localization at $p$ does not change a $p$-local spectrum.
\end{itemize}
\end{remark}  

\begin{proposition}\label{prob:GEM}
    $\operatorname{ cofib}( \operatorname{L}^s(\IZ)\rightarrow 
 \operatorname{L}^s(\IQ))$ is a generalized Eilenberg-MacLane spectrum.
\end{proposition}
\begin{proof}
    It suffices to check that the $\operatorname{ cofib}( \operatorname{L}^s(\IZ)\rightarrow 
 \operatorname{L}^s(\IQ))$ is an $H\bZ$ -- module. Our proof uses the arguments used for the proofs of Theorem A of \cite{TaylorWilliams} and Theorem 1 of \cite{land2022remark}. The former reference concerns the non-genuine symmetric $L$-theory, while the latter concerns the genuine $L$-theory used in this paper. First of all, $\L^{s}(\IZ)$ is an  $\mathbb E_{\infty}$-ring and, for any $R$, $\L^s(R)$ is a module over $\L^s(\IZ)$ \cite{Hebestreit_2020,land2022remark}.  Also, the spectrum level cofiber of a map of $\L^s(\bZ)$-modules has a natural structure of a $\L^s(\bZ)$-module (Theorem 1.1, Chapter III, \cite{elmendorf1997rings}). 
    
Consider the composition $H\IZ\xrightarrow{M\eta_{(2)}} \operatorname{MSO}_{(2)}\xrightarrow{\sigma^s_{(2)}} \tau_{\geqslant 0}\L^{s}(\IZ)_{(2)}$. Here, $M\eta_{(2)}$ is a map of $\mathbb E_2$-rings defined in
Theorem 3.6 of \cite{Hebestreit_2020} and $\sigma^s_{(2)}$ is
Ranicki's symmetric signature $\sigma^s$, see, e.g.,
\cite{ranicki1992algebraic}, localized at $2$. The symmetric signature is a map of $\mathbb E_{\infty}$-rings \cite{land2022remark}. 

The symmetric signature $\sigma^s:MSO\to \tau_{\geqslant 0}\L^s(\IZ)$ is a map of connective spectra, while we can extend it to a map $\hat\sigma^s:MSO\to \L^s(\IZ)$ as follows. Since an $(\infty,1)$-functor $X\mapsto \tau_{\geqslant 0}X$, which maps a spectrum $X$ to its connective truncation, is right-adjoint, the counit of the adjunction gives the canonical map $\tau_{\geqslant 0}X\to X$. Note that the smash product of two connective spectra is still connective, so the embedding functor is symmetric monoidal and, by \cite{lurie2017higher}, the counit of the adjunction is a map of $\mathbb E_\infty$-rings.

 By the localization exact sequence and the calculations in
Section~\ref{sec:L-groups}, all homotopy groups of
$\operatorname{cofib}
    \bigl(\L^s(\IZ)\longrightarrow\L^s(\IQ)\bigr)$
are $2$-primary. Hence this spectrum is $2$-local, and
\begin{align}
    \operatorname{cofib}
    \bigl(\L^s(\IZ)\longrightarrow\L^s(\IQ)\bigr)
    \cong
    \operatorname{cofib}
    \bigl(\L^s(\IZ)\longrightarrow\L^s(\IQ)\bigr)_{(2)}.
\end{align}

 We obtain that $\operatorname{ cofib}( \L^s(\IZ)\rightarrow 
 \L^s(\IQ))$ is an $\L^{s}(\IZ)_{(2)}$-module and, consequently, an $H\IZ$-module. 
\end{proof}
\vspace{3mm}

We define the spectrum of invertible stabilizer codes
\begin{align}\label{eqref:spec_inv}
\operatorname{ISC}\coloneqq \tau_{\leqslant 0}\Sigma \operatorname{ cofib}( \L^s(\IZ)\rightarrow   
 \L^s(\IQ))\,.
\end{align}
So, $\pi_{-\dd}(\operatorname{ISC})= \operatorname{ISC}_{\dd}$ for $\dd\geqslant 0$ and vanishing otherwise, where $\operatorname{ISC}_{\dd}$ are determined in \eqref{eq:CliffordPhases}.
By Proposition \ref{prob:GEM}, we obtain non-canonically $\operatorname{ISC}$ as a plain spectrum:
 \begin{align}
\operatorname{ISC}\cong \bigvee_{k< 0}\Sigma ^{4k+1}H\CW^{pt}\bigvee_{k<0}\Sigma ^{4k} H\IZ/2\,.
\end{align}

\subsection{Relation with Kitaev's conjecture}
 It is a famous conjecture by Alexei Kitaev \cite{Kitaev} that there exists a generalized cohomology theory corresponding to invertible states, see \cite{kapustin2014symmetryprotectedtopologicalphases,xiong2018minimalist,Gaiotto_2019,Freed_2021,Shiozaki2022}. There has been recent progress on this conjecture \cite{kubota2025stablehomotopytheoryinvertible}, while the determination of the spectrum of the invertible states is still an open problem. Here we have proposed a classification of invertible \textit{stabilizer} phases which a priori is different from the classification of the general invertible phases. Our classification scheme led us to the homotopy groups of the relative $L$-theory spectrum \eqref{eqref:spec_inv}, which is a generalized Eilenberg-MacLane spectrum by Proposition \ref{prob:GEM}. Scalar phases are present in the operator groups, but the classification retains only the phase-free label data and does not keep the central $\operatorname{U}(1)$ as an independent invariant. For this reason our spectrum does not have a $\operatorname{U}(1)$ among its homotopy groups; compare this with an alternative construction of \cite{czajka2025anomalieslatticehomotopyquantum}.
\vspace{5mm}

\textbf{Acknowledgements.}
The authors thank Theo Johnson-Freyd, Jeongwan Haah, Marcus Land, Jens Hornbostel, and Chuck Weibel for fruitful correspondence, and Nathanan Tantivasadakarn and Ryan Thorngren for helpful discussions. RG thanks Yichen Hu and Bowen Yang for collaboration on related projects. RG acknowledges support by the Mani L. Bhaumik Institute for Theoretical Physics. RG acknowledges support from the Simons Center for Geometry and Physics, Stony Brook University at which some of the research for this paper was performed.

\begin{appendices}
    \section{Majorana stabilizer codes}\label{app:majorana}
In this appendix we comment on potential applications of odd anti-hermitian forms to Majorana stabilizer codes. The Majorana group is a group within the $\IZ/2$-graded algebra of operators acting on a $\IZ/2$-graded Hilbert space  $(\IC^{1|1})^{\hat \otimes n}$ \cite{Bravyi_2002,bettaque2025structuremajoranacliffordgroup}, where $\hat \otimes$ is the $\IZ/2$-graded tensor product. We define the Majorana operators by the following relations:
\begin{align}
    \chi_i\chi_j+\chi_j\chi_i=2\delta_{i,j}\,,\quad\quad \chi^{*}_i=\chi_i\,.
\end{align} 
The Majorana group is the  group generated by the Majorana operators and phases: 
\begin{align}
    \CM_{2n}=\la \omega \,\chi_1^{x_1}\ldots \chi_{2n}^{x_{2n}}\,|\,\omega \in U(1)\,, x_1,\ldots, x_{2n}\in \IZ/2\ra\,.
\end{align}
We label Majorana strings up to phases with the elements of $(\IZ/2)^{2n}$
\begin{align}
(\IZ/2)^{2n}\ni x=(x_1,\ldots, x_{2n})\mapsto \mu(x)\coloneqq \chi_1^{x_1}\ldots \chi_{2n}^{x_{2n}}
\end{align}
and express the commutator as follows
\begin{align}
    [\omega_1 \mu(x),\omega_2\mu(x')]=e^{2\pi i \kappa(x,x')}
\end{align}
where $\kappa$ is a $\frac{1}{2}\IZ/\IZ$-valued anti-symmetric form on $(\IZ/2)^{2n}$ with the following matrix  
    \begin{align}\label{Maj_commutator}
         \begin{pmatrix}
            0 & 1/2 & \ldots & \ldots &  1/2\\
            1/2& 0 & 1/2 & \ldots & \vdots \\
            \vdots&  & & & \vdots\\
            \vdots&  & & & 1/2\\
            1/2 &\ldots & \ldots & 1/2 & 0
        \end{pmatrix}\,.
    \end{align}

The Majorana group is $\IZ/2$-graded by the grading operator $P$ such
that $P^*=P$, $P^2=1$ and $P\chi_iP=-\chi_i$. Define the parity
homomorphism
\begin{align}\label{eq:z2grading}
    c:(\IZ/2)^{2n}
    \longrightarrow
    \frac{1}{2}\IZ/\IZ,
    \qquad
    c(x)
    \coloneqq
    \frac{x_1+\cdots+x_{2n}}{2}.
\end{align}
Then
\begin{align}\label{eq:parity}
    P\mu(x)P
    =
    e^{2\pi i c(x)}\mu(x).
\end{align}
The elements of $\CM_{2n}$ of even parity form a subgroup which we
denote by $\CM_{2n}^+$. A \textit{Majorana stabilizer code} is an
isotropic subgroup
\[
    F\subseteq(\IZ/2)^{2n}
\]
for the form $\kappa$ in \eqref{Maj_commutator} such that
$F\subseteq\ker c$
\cite{Bravyi_2002,bettaque2025structuremajoranacliffordgroup}.

Let us connect the description of the Majorana group using
\eqref{Maj_commutator} with odd anti-symmetric linking forms over
$(\IZ,\IZ-\{0\})$. We define a form $F_2$ on
$\CD_2=\IZ/2\oplus\IZ/2$ with the following matrix
\begin{align}\label{eq:Fform}
    \begin{pmatrix}
        0 & 1/2\\
        1/2 & 1/2
    \end{pmatrix}.
\end{align}
Let $(\CD,b)=nF_2$. In each copy of $F_2$, set
\[
    f_1=e_2,
    \qquad
    f_2=e_1+e_2.
\]
In the resulting basis $\{f_i\}_{i=1}^{2n}$,
\begin{align}
    \hat b(f_i,f_j)
    =
    \frac{1}{2}\delta_{i,j}.
\end{align}
Under the identification of $\CD$ with $(\IZ/2)^{2n}$ given by this
basis, the characteristic homomorphism
\[
    x\longmapsto\hat b(x,x)
\]
is the parity homomorphism $c$. Let
$\bar c:\CD\to\IZ/2$ be defined by
$c(x)=\bar c(x)/2$, and define
\begin{align}
    \kappa_b(x,y)
    \coloneqq
    \hat b(x,y)
    +
    \frac{\bar c(x)\bar c(y)}{2}.
\end{align}
The form $\kappa_b$ is alternating, and in the basis $\{f_i\}$ its
matrix is exactly \eqref{Maj_commutator}. Thus, after identifying this
basis with the standard Majorana labels, the commutation relation is
\begin{align}\label{eq:modified_commutation}
    \mu(x)\mu(y)
    =
    \mu(y)\mu(x)e^{2\pi i\kappa_b(x,y)}.
\end{align}
We note that there are odd anti-symmetric forms on finite abelian
groups that are not isomorphic to direct sums of copies of $F_2$;
they do not correspond to the ordinary Majorana groups $\CM_{2n}$.

\end{appendices}
\bibliography{InvStates}
\end{document}